\documentclass[12pt,aps,tightenlines,amsmath,amssymb,nofootinbib,titlepage,preprintnumbers,superscriptaddress]{revtex4-1}
\usepackage[utf8]{inputenc}
\usepackage{amsmath}
\usepackage{amssymb}
\usepackage{graphicx}
\usepackage{xcolor}
\usepackage{hyperref}

\usepackage{mciteplus}

\mciteErrorOnUnknownfalse

\begin{document}
\preprint{FERMILAB-PUB-25-0154-PPD}

\title{
\Large {The Short-Baseline Near Detector at Fermilab} \\ 
\large{Input to the European Strategy for Particle Physics 2026 Update}}

\collaboration{SBND Collaboration}\thanks{Spokespeople: Ornella Palamara (palamara@fnal.gov) and David Schmitz (dwschmitz@uchicago.edu)}


\author{R.~Acciarri}\affiliation{Fermi National Accelerator Laboratory, Batavia, Illinois 60510, USA}
\author{L.~Aliaga-Soplin}\affiliation{University of Texas at Arlington, TX 76019, USA}
\author{O.~Alterkait}\affiliation{Tufts University, Medford, MA, 02155, USA}
\author{R.~Alvarez-Garrote}\affiliation{CIEMAT, Centro de Investigaciones Energ\'{e}ticas, Medioambientales y Tecnol\'{o}gicas, Madrid E-28040, Spain}
\author{D.~Andrade Aldana}\affiliation{Illinois Institute of Technology, Chicago, IL 60616, USA}
\author{C.~Andreopoulos}\affiliation{University of Liverpool, Liverpool L69 7ZE, United Kingdom}
\author{A.~Antonakis}\affiliation{University of California, Santa Barbara CA, 93106, USA}
\author{L.~Arellano}\affiliation{University of Manchester, Manchester M13 9PL, United Kingdom}
\author{W.~Badgett}\affiliation{Fermi National Accelerator Laboratory, Batavia, Illinois 60510, USA}
\author{S.~Balasubramanian}\affiliation{Mount Holyoke College, South Hadley, MA 01075, USA}
% ...

\clearpage

\begin{abstract}
\vspace{40mm}
{\centering {\bf\large ~~~~~~~~~~~~~~~~~~~~~ Executive Summary}}\\ \\ \\
The Short-Baseline Near Detector (SBND) is a 112-ton liquid argon time projection chamber (LArTPC) neutrino detector located 110 meters from the Booster Neutrino Beam (BNB) target at Fermilab. Its main goals include searches for eV-scale sterile neutrinos as part of the Short-Baseline Neutrino (SBN) program, other searches for physics beyond the Standard Model, and precision studies of neutrino-argon interactions. In addition, SBND is providing a platform for LArTPC neutrino detector technology development and is an excellent training ground for the international group of scientists and engineers working towards the upcoming flagship Deep Underground Neutrino Experiment (DUNE).

SBND began operation in July 2024, and started collecting stable neutrino beam data in December 2024 with an unprecedented rate of ${\sim} 7,000$ neutrino events per day. During its currently approved operation plans ($2024-2027$), SBND is expected to accumulate an exposure of around $10 \times 10^{20}$ protons on target, recording nearly 10 million neutrino interactions. The near detector dataset will be instrumental in testing the sterile neutrino hypothesis with unprecedented sensitivity in SBN and in probing signals of beyond the Standard Model physics. It will also be used to significantly advance our understanding of the physics of neutrino-argon interactions ahead of DUNE. After the planned accelerator restart at Fermilab (2029+), opportunities are being explored to operate SBND in antineutrino mode in order to address the scarcity of antineutrino–argon scattering data, or in a dedicated beam-dump mode to significantly enhance sensitivity to searches for new physics.

SBND is an international effort, with approximately $40\%$ of institutions from Europe, contributing to detector construction, commissioning, software development, and data analysis. Continued European involvement and leadership are essential during SBND's operations and analysis phase for both the success of SBND/SBN and its role leading up to DUNE.

\end{abstract}

\maketitle

\tableofcontents

\section{Introduction and Current Status}\label{sec:introduction}

The Short-Baseline Near Detector (SBND) is a 112-ton liquid argon time projection chamber (LArTPC), positioned 110 meters from the Booster Neutrino Beam (BNB) target, and serves as the near detector of the Short-Baseline Neutrino (SBN) program at Fermilab~\cite{MicroBooNE:2015bmn}. Due to the large detector mass and location close to the target, the science of SBND includes a rich program as part of the SBN Program and on its own, addressing explanations of the short-baseline neutrino anomalies~\cite{Machado:2019oxb}, searching for physics beyond the Standard Model, and conducting precision studies of neutrino-argon interactions. The SBN Program is fully online now with both the near (SBND) and far (ICARUS) detectors operating and is poised to test the eV-scale sterile neutrino hypothesis by covering the parameter regions allowed by past anomalies at ${\sim} 5\sigma$ significance. 
As the near detector in the SBN Program, SBND plays a key role in mitigating large neutrino flux and cross section uncertainties.

The SBN Program aligns closely with recommendations from major particle physics roadmaps. The 2014 US Particle Physics Project Prioritization Panel (P5) recommended a world-leading short-baseline neutrino program at Fermilab with strong domestic and international participation, particularly in preparation for the Long-Baseline Neutrino Facility (LBNF) and the Deep Underground Neutrino Experiment (DUNE)~\cite{DUNE:2020lwj}. The 2023 P5 report~\cite{P5:2023wyd} reaffirmed this vision, highlighting SBN’s role in conclusively testing short-baseline neutrino anomalies, advancing the LArTPC technology, and refining neutrino-argon interaction measurements. Additionally, European strategy reports, such as the previous Physics Briefing Book~\cite{EuropeanStrategyforParticlePhysicsPreparatoryGroup:2019qin} and the 2020 Update of the European Strategy for Particle Physics~\cite{CERN-ESU-015}, have recognized SBND and ICARUS as uniquely positioned to address these anomalies and emphasized the importance of ongoing European collaboration in US neutrino experiments leading to DUNE.

SBND is an international collaboration with significant contributions from European institutions, including CERN, Spain, Switzerland, and the UK. Approximately 40\% of SBND-affiliated universities and research institutions are based in Europe, and they have made critical contributions to the SBND design, construction, software development, detector operation and physics exploitation. These European contributions, largely supported by national funding agencies and European funds, play a crucial role in both SBND and the broader development towards DUNE. 

\begin{figure}
    \centering
    \includegraphics[width=0.48\linewidth]{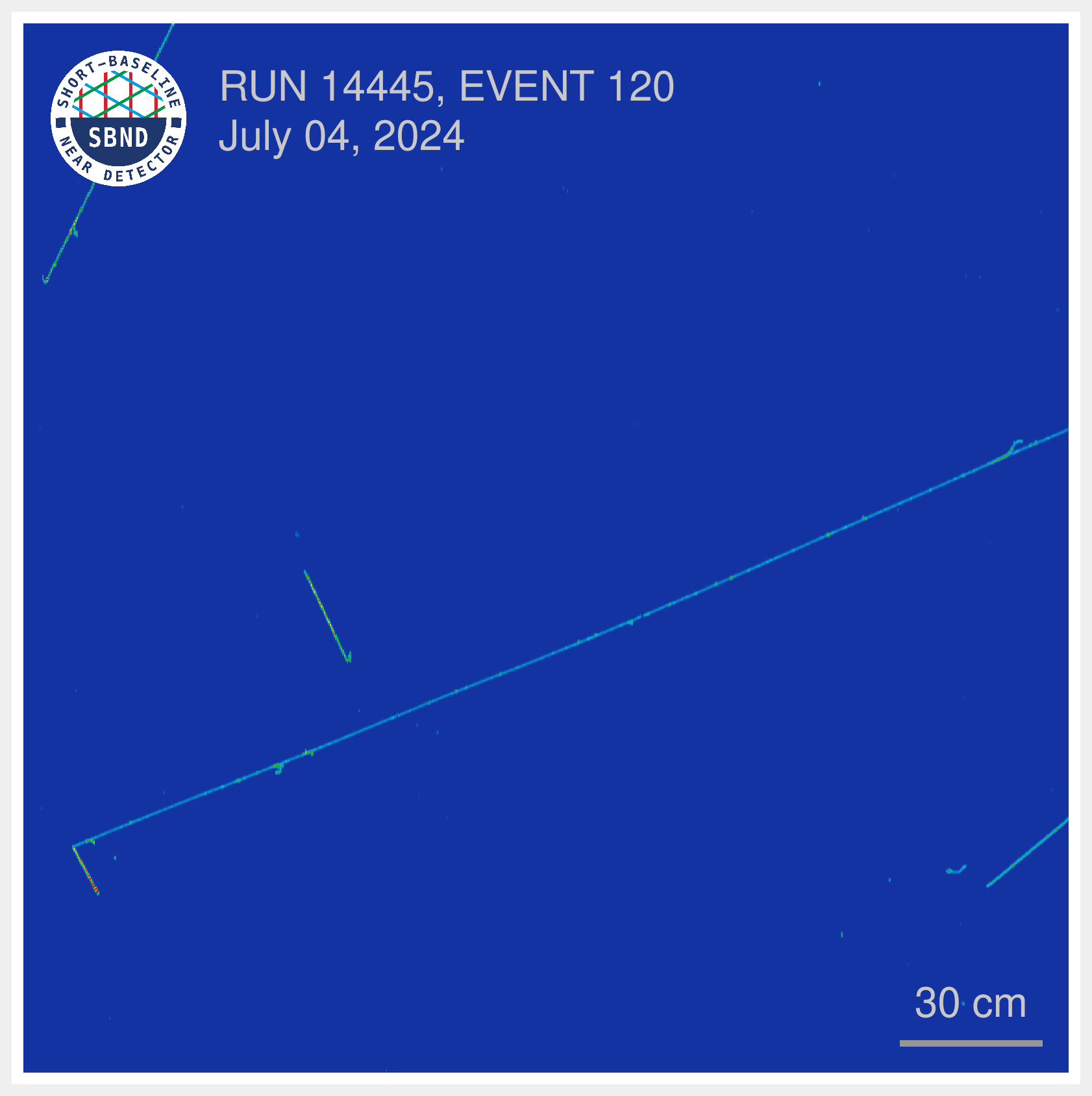}
    \includegraphics[width=0.48\linewidth]{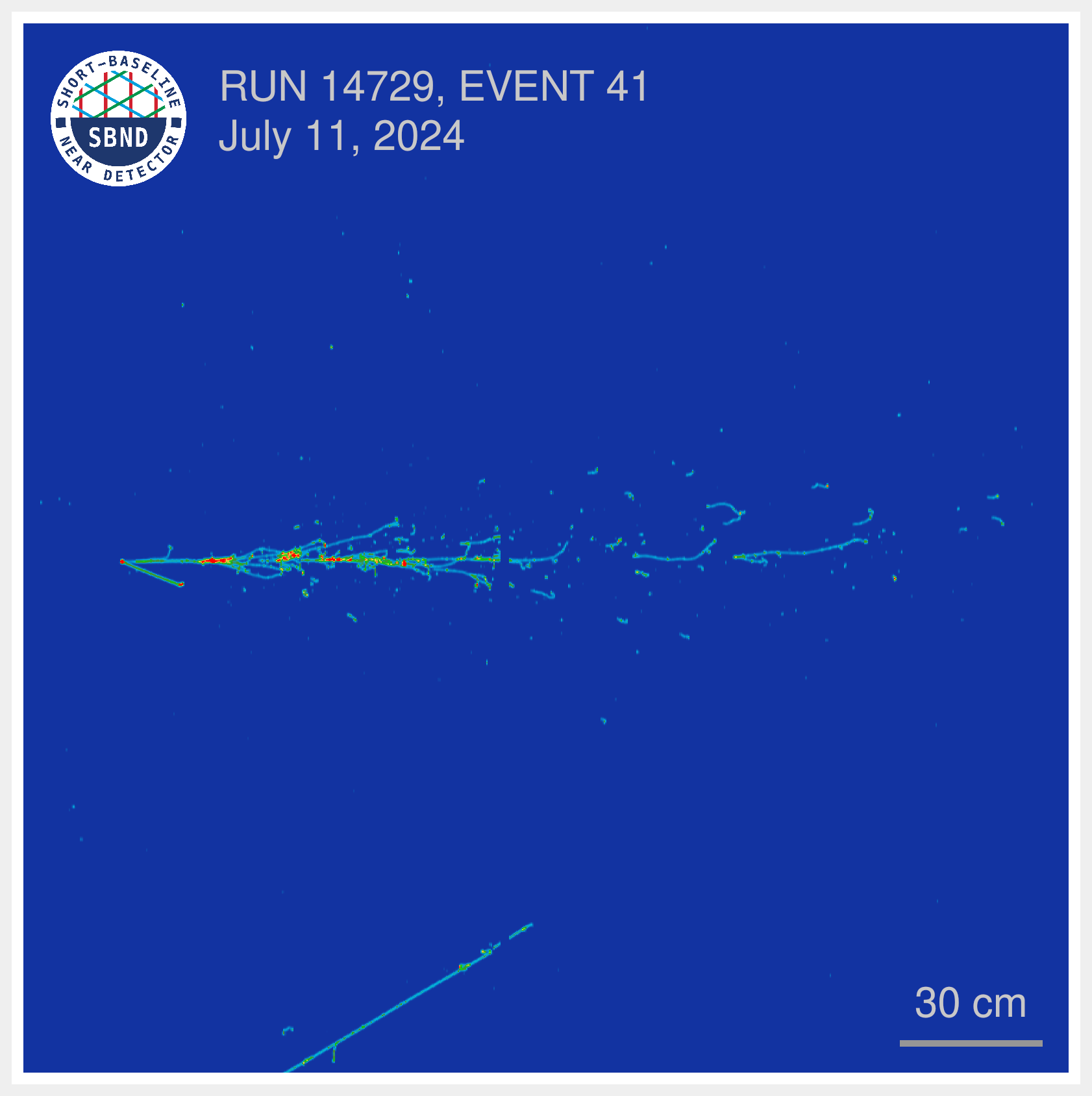}
    \caption{Sample $\nu_\mu$ CC (left) and  $\nu_e$ CC (right) candidate events observed in the SBND data.}
    \label{fig:neutrino_events}
\end{figure}

After nearly a decade of design, construction, and installation, SBND began operation in July 2024 and started collecting stable BNB data in December 2024 with an unprecedented rate of ${\sim} 7,000$ neutrino events per day. In Fig.~\ref{fig:neutrino_events}, we show two of the first $\nu_\mu$ and $\nu_e$ charged-current (CC) candidate events observed in the SBND data. SBND has already accumulated the world’s largest neutrino-argon interaction dataset and is expected to continue to operate until 
the planned long accelerator shutdown at Fermilab (scheduled for late 2027/early 2028), accruing a total exposure of $10 \times 10^{20}$ protons on target.
As this white paper outlines, continued support for SBND’s operations and analysis phase, as well as future planning, is critical. This support is necessary not only for realizing the full scientific potential of the SBN Program and SBND but also for informing the next-generation flagship experiment, DUNE.

\section{Physics Goals and Capabilities}\label{sec:physics_goals}

Utilizing the LArTPC technology, which offers exceptional particle identification and fine-sampling calorimetry, the SBND physics reach benefits from its large mass (112~tons) and close proximity (110~m) to a high-intensity beam. Beyond the eV mass-scale sterile neutrino search as part of the SBN Program, SBND has a rich physics program probing beyond the Standard Model physics, and conducting precision studies of neutrino-argon interactions. 

\subsection{Search for eV-Scale Sterile Neutrinos as Part of the SBN Program}

\begin{figure}
    \centering
    \includegraphics[width=0.49\linewidth]{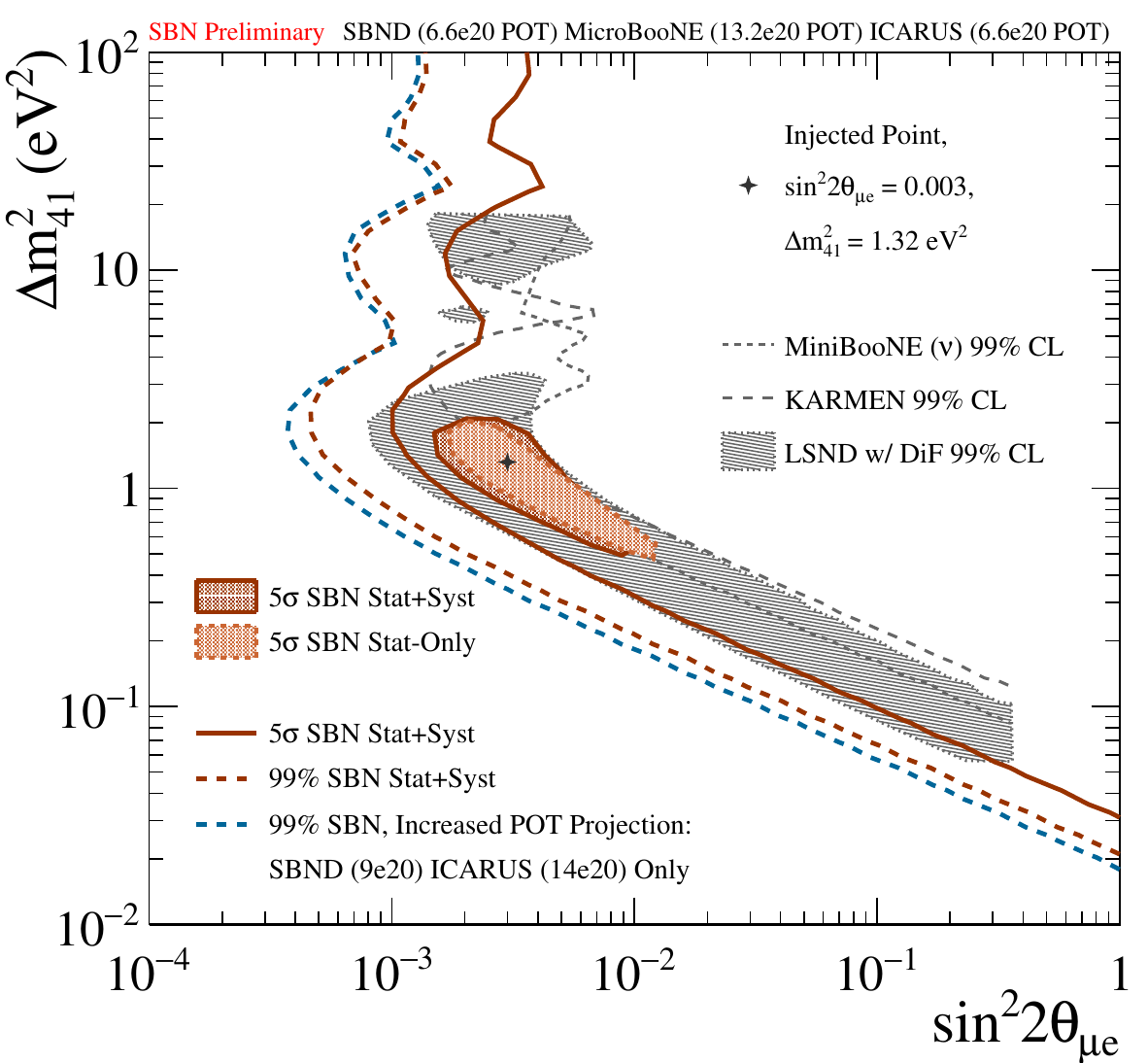}
    \includegraphics[width=0.49\linewidth]{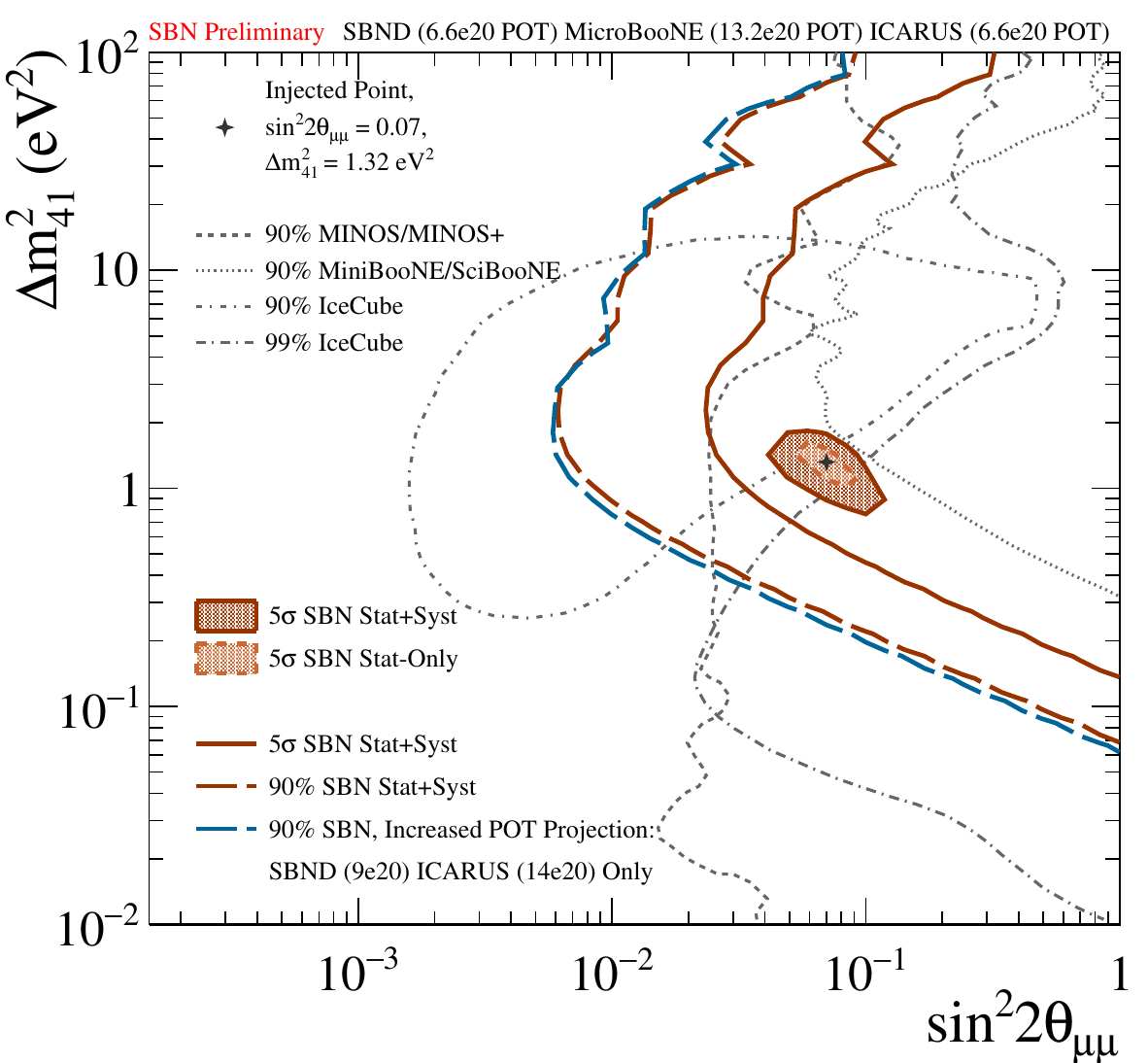}
    \caption{
    The SBN exclusion and allowed-region sensitivities to $\nu_e$ appearance (left) and $\nu_\mu$ dis\-ap\-pear\-ance (right) under the (3+1) sterile neutrino hypothesis. External contours from Refs.~\cite{Dentler:2018sju,SciBooNE:2011qyf,MINOS:2017cae,IceCube:2020tka}.}
    \label{fig:neutrino_osc_sensitivities}
\end{figure}

As the near detector of the SBN Program, SBND will perform a high-precision measurement of the BNB neutrino flux and the neutrino-argon cross-section close to the source. The use of the same target and detector technology in all SBN detectors grants a high correlation between the near and far detectors, reducing systematic uncertainties. This increases the sensitivity for oscillations at $\Delta m^2 \sim 1~\rm{eV}^2$ and enables the SBN Program to conclusively address the short-baseline neutrino oscillation anomalies, as shown in Fig.~\ref{fig:neutrino_osc_sensitivities} for exclusive appearance and disappearance channel searches. A multi-channel oscillation search for sterile neutrinos is essential to confirm or exclude any oscillation signal. The observation of $\nu_e$ appearance, which in the (3+1) sterile neutrino hypothesis is driven by $\sin^2 2\theta_{\mu e} = 4|U_{\mu 4}|^2|U_{e 4}|^2$, must be accompanied by $\nu_\mu$ disappearance proportional to $\sin^2 2\theta_{\mu \mu} = 4|U_{\mu 4}|^2|(1-|U_{\mu 4}|^2)$.
In addition, the SBN Program can perform a flavor-inclusive neutral current disappearance search as a function of the baseline~\cite{Furmanski:2020smg}.

\subsection{Precision Measurements of Neutrino-Argon Scattering Cross Sections}

\begin{figure}
    \centering
    \includegraphics[width=0.32\linewidth]{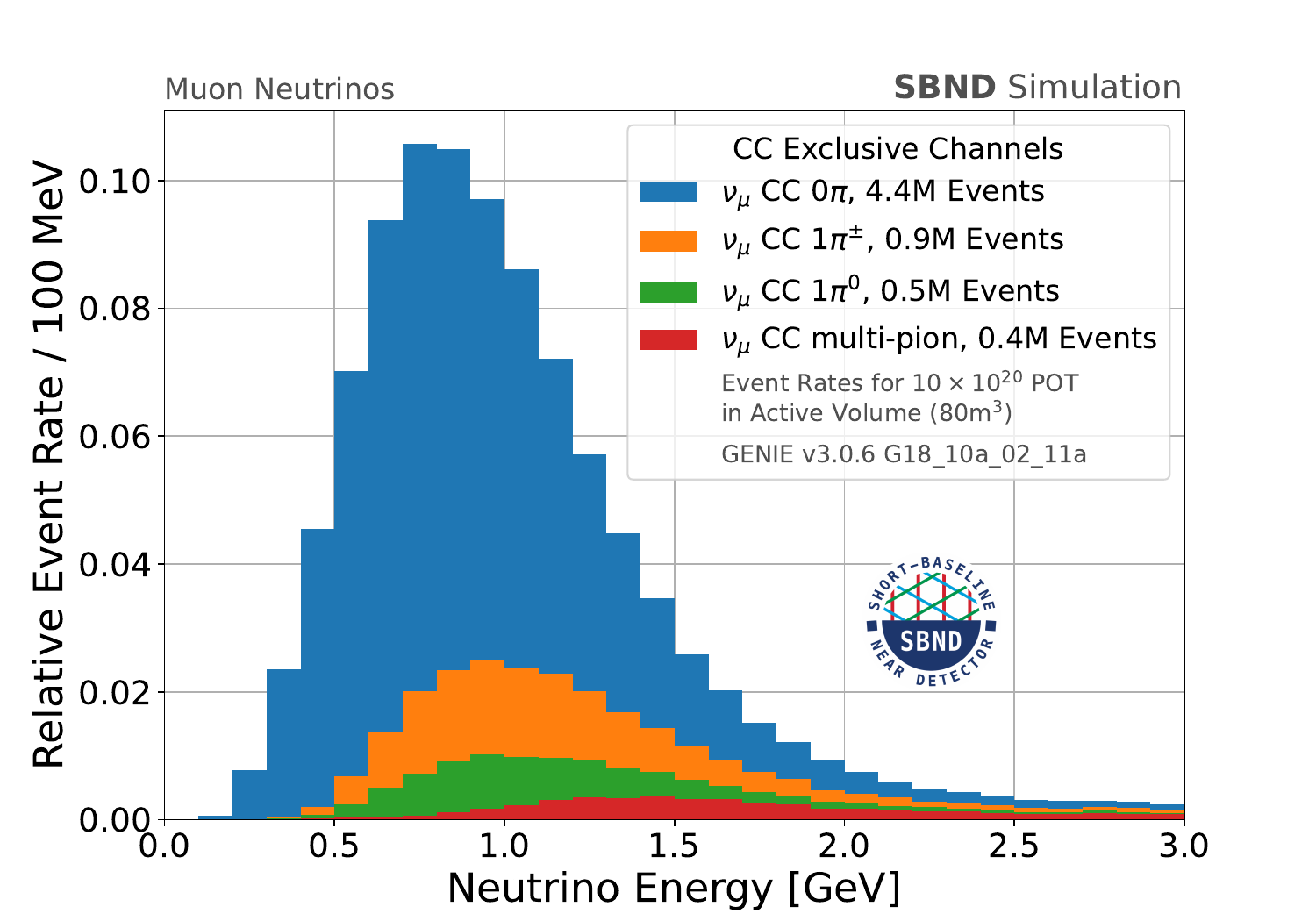}
    \includegraphics[width=0.32\linewidth]{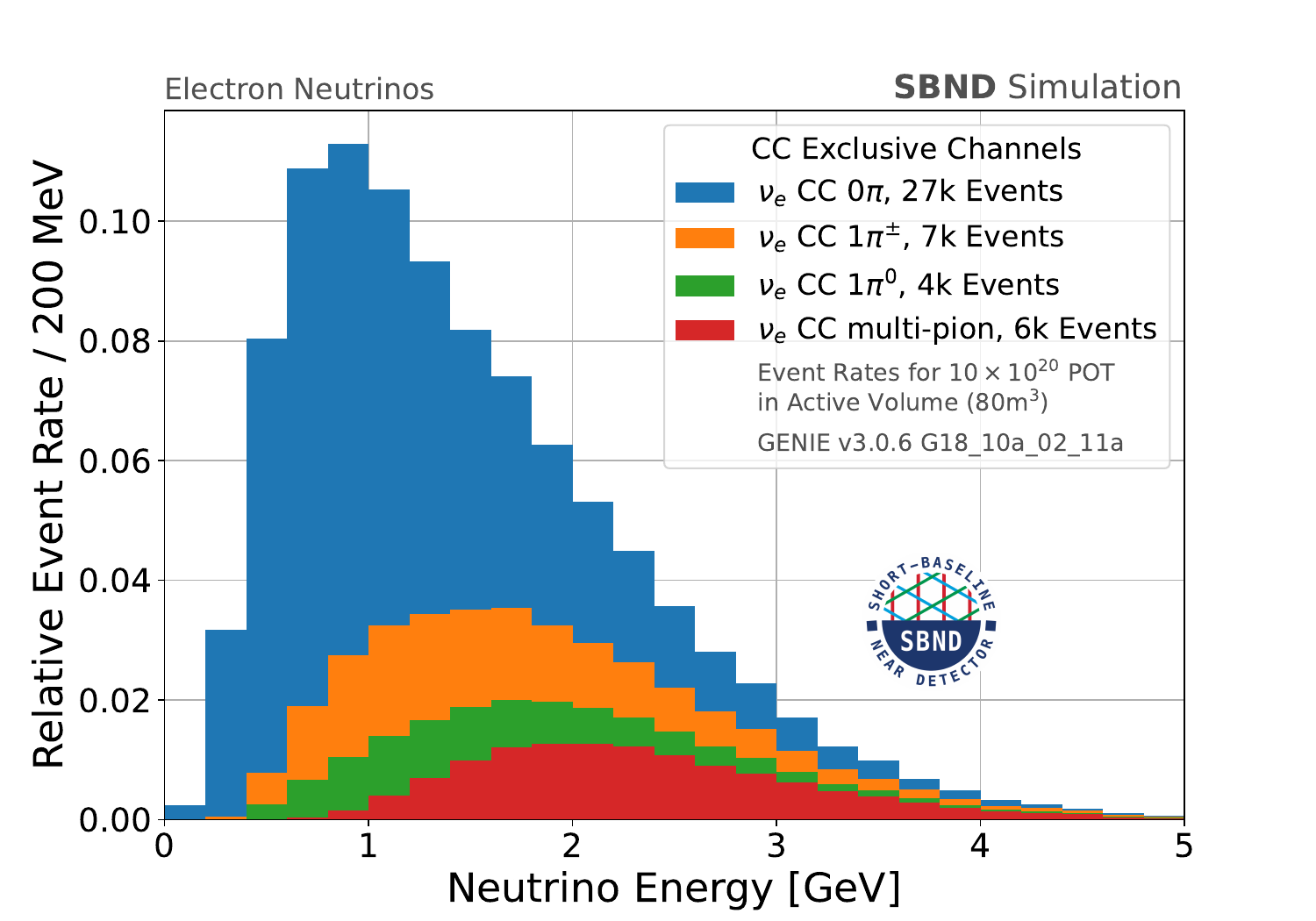}
    \includegraphics[width=0.33\linewidth]{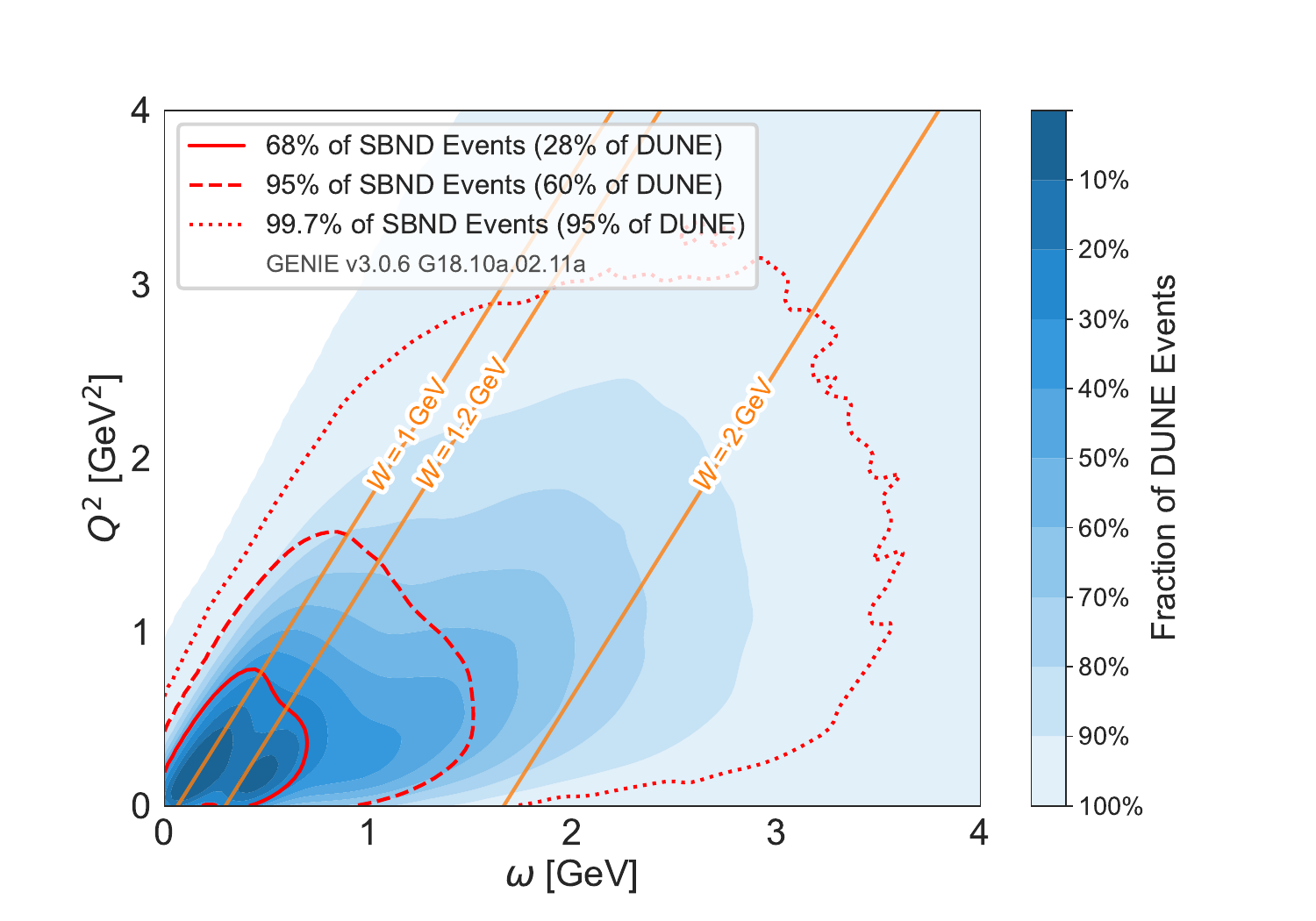}
    \caption{The expected SBND $\nu_\mu$ CC (left) and  $\nu_e$ CC (middle) event rates for $10 \times 10^{20}$ protons on target (POT) exposure in forward horn current (neutrino) mode of the BNB. This is approximately the POT expected from the ongoing three-year run between 2025--2027. Right: SBND $\nu_\mu$ CC kinematic coverage overlaid on the DUNE $\nu_\mu$ CC kinematic coverage.}
    \label{fig:neutrino_event_rate}
\end{figure}

Understanding neutrino-nucleus interactions is critical to the success of neutrino oscillation experiments, including the SBN Program and DUNE that share argon as the target nucleus~\cite{Balantekin:2022jrq}. Due to its proximity to the neutrino source, SBND will collect an unprecedentedly high event rate of neutrino interactions, providing an ideal avenue for precision studies of neutrino-argon interactions in the sub-GeV and GeV energy range. SBND will record approximately ${\sim} 7,000$ neutrino events per day, amounting to ${\sim} 2,000,000$ $\nu_\mu$ CC events per year and ${\sim} 15,000$ $\nu_e$ CC events per year, enabling both inclusive and exclusive neutrino-argon interaction measurements. Fig.~\ref{fig:neutrino_event_rate} shows the spectra of $\nu_\mu$ CC (left) and $\nu_e$ CC (middle) expected event rates for an exposure of $10 \times 10^{20}$ protons on target. Before the DUNE era, SBND will produce the world's highest-statistics neutrino-argon cross-section measurements. To place the SBND measurements into the context of other liquid argon detectors, each year of exposure of SBND will provide an event sample a few times larger than the
one available from the full MicroBooNE operation. 

Note that the BNB neutrino flux spectrum at SBND peaks near the neutrino energy of the second oscillation maximum for DUNE (${\sim} 0.8$~GeV) and includes a substantial sample up to the first DUNE oscillation maximum (${\sim} 2.6$~GeV). Fig.~\ref{fig:neutrino_event_rate} (right) demonstrates that the SBND kinematic phase space has substantial overlap with the DUNE kinematic phase space. Leveraging the high resolution of LArTPC technology, SBND will perform precise measurements of multiple final states for both $\nu_\mu$ and $\nu_e$ events. Furthermore, the high event rate will also enable the detection of several thousands of events in rare interaction channels, such as the production of hyperons $\Lambda^{0}$ and $\Sigma^{+}$ in neutrino-argon scatterings. Until the planned long accelerator shutdown at Fermilab,
SBND is expected to yield nearly 10 million neutrino interactions on argon across charged-current and neutral-current channels. This dataset will be instrumental in advancing our understanding of neutrino-argon scattering at the sub-GeV and GeV energy scales before the DUNE era.

\subsection{Other Searches for Physics Beyond the Standard Model}

The close location of SBND to the neutrino beam origin enables other searches for Beyond-Standard-Model (BSM) particles produced in meson decays, particularly from pions and kaons, as well as in proton-target interactions. As an intensity-frontier experiment, SBND will be able to explore a complementary phase space to the energy-frontier experiments, restricted to searching for new particles with lower masses but reaching smaller couplings. SBND is expected to have one of the leading sensitivities to muon-coupled heavy neutral lepton searches below the kaon mass until the DUNE Near Detector turns on~\cite{Ballett:2016opr, Ballett:2019bgd}. In addition, SBND has competitive sensitivity to other dark sector particles such as a Higgs-Portal scalar~\cite{Batell:2019nwo}, heavy axion-like particles~\cite{Berger:2024xqk, Kelly:2020dda} or dark photons that mediate interactions with light dark matter~\cite{deNiverville:2016rqh, Batell:2021ooj}.

Going beyond sterile neutrino oscillations, SBND will test a novel class of models predicting new physics in meson decays~\cite{Dutta:2021cip} or dark sectors accessible via neutrino interactions~\cite{Abdullahi:2023ejc} to explain the MiniBooNE excess. Moreover, there are ongoing efforts to carry out searches using more model-independent approaches, such as using simplified frameworks featuring generic long-lived particles~\cite{Batell:2023mdn}.

The SBND Collaboration is working with the theory community to realize all these searches. Similar to the collider community, where a theory-experiment collaboration has successfully delivered established tools to simulate BSM physics in collider detectors, the SBN Program is a unique platform for the accelerator-neutrino community to develop BSM simulation tools that interface with neutrino beamline simulations and detector simulations, developing infrastructure that will also benefit future searches in the DUNE experiment.

In order to maximize the sensitivity of these searches, SBND will have to reduce the SM neutrino background. In addition to the excellent capabilities of the LArTPC technology to reconstruct exclusive final states, SBND is developing new techniques such as an advanced timing reconstruction based on scintillation light that enables the separation of the massive long-lived particles from the SM neutrinos based on their time of flight~\cite{SBND:2024vgn}.

\section{Features of the Beam and Detector}\label{sec:beam_detector}

\subsection{BNB and SBND-PRISM Concept}

The SBN Program utilizes the Booster Neutrino Beam (BNB) at Fermilab. This neutrino beam is produced by extracting protons with a kinetic energy of 8 GeV from the Booster accelerator and directing them onto a beryllium target, generating a secondary beam of hadrons. The charged secondaries are then focused by a single toroidal aluminum alloy focusing horn surrounding the target. The BNB has been operating successfully and stably for more than two decades. In neutrino mode, where positively charged hadrons are focused, the flux composition is energy-dependent but predominantly consists of muon neutrinos ($\nu_\mu$), making up ${\sim} 92.5\%$ of the total flux. This is followed by muon antineutrinos ($\bar{\nu}_\mu$) at ${\sim} 6.9\%$, with an intrinsic contamination from electron neutrinos and antineutrinos ($\nu_e/\bar{\nu}_e$) at approximately ${\sim} 0.6\%$.

Constraining BNB flux uncertainties has greatly benefited from dedicated hadron production data collected by the HARP experiment at CERN~\cite{HARP:2007dqt}. These data have played a crucial role in reducing uncertainties by constraining the primary source of flux systematic uncertainty: pion production in proton-beryllium interactions. Despite this improvement, the total flux uncertainty -- which also includes contributions from other sources, such as interactions occurring outside HARP’s phase space coverage -- remains approximately 8\% at the peak of the muon neutrino flux and increases in both the low- and high-energy regions~\cite{MiniBooNE:2008hfu}. New data~\cite{Nagai:2783037} with broader phase space coverage and reduced uncertainties will be valuable for SBND in further constraining the flux uncertainty. Dedicated measurements in the BNB energy region, such as those proposed for the new tertiary low-energy beam from the NA61/SHINE experiment at CERN, will be crucial. This setup will allow us to reduce uncertainties not only in primary beryllium interactions but also in secondary interactions in beryllium and other materials, ultimately improving the precision of flux predictions. In particular, these improvements will strengthen constraints on the low-energy and high-energy falling slopes of the BNB flux, refining neutrino flux predictions and enhancing the robustness of SBN and SBND analyses. The potential impact of new NA61 data on reducing SBND’s flux uncertainty has been estimated under reasonably attainable assumptions regarding the production of charged pions and kaons in proton-beryllium interactions, based on published NA61 data. 

\begin{figure}
    \centering
    \includegraphics[width=0.50\linewidth]{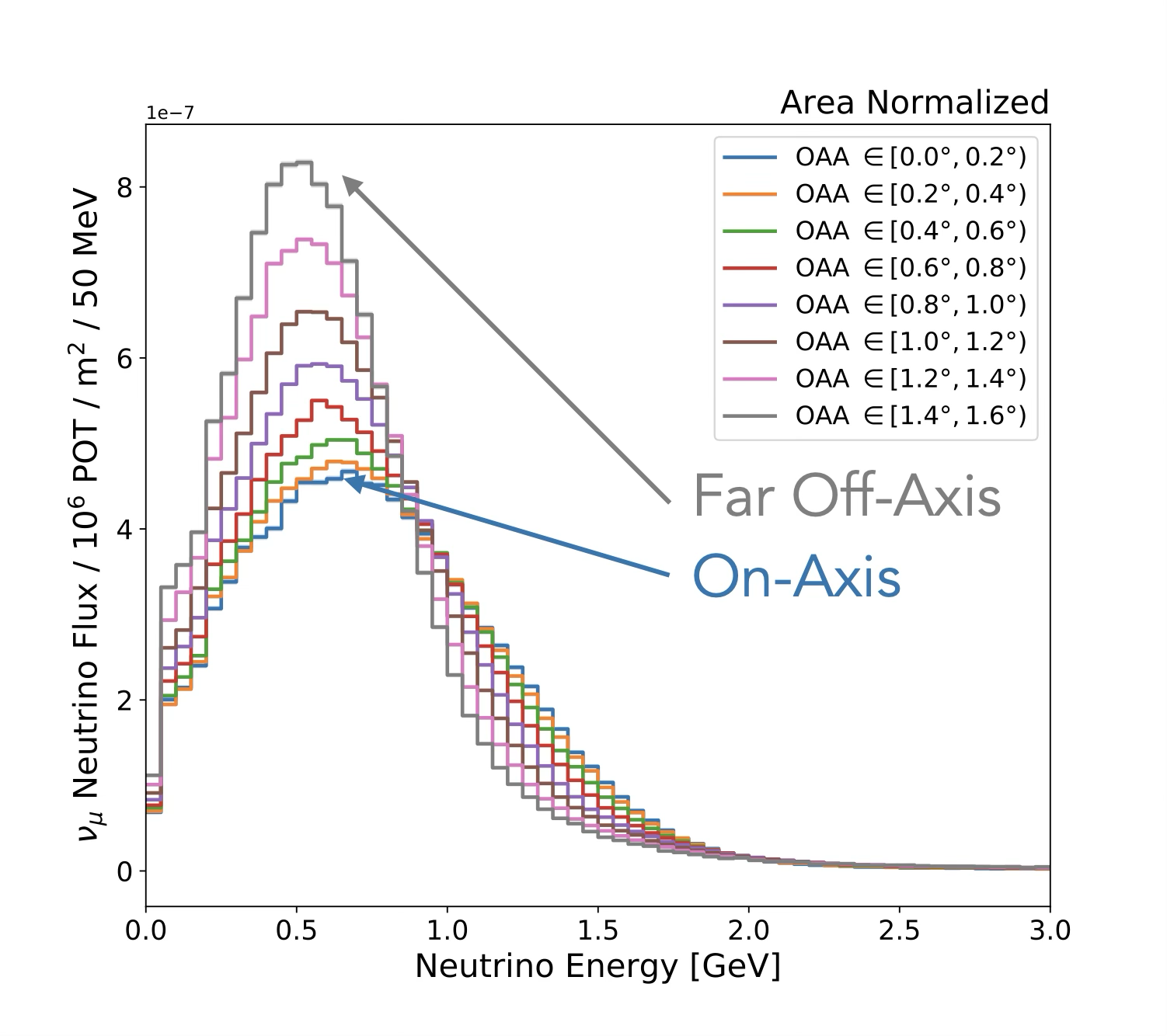}
    \includegraphics[width=0.46\linewidth]{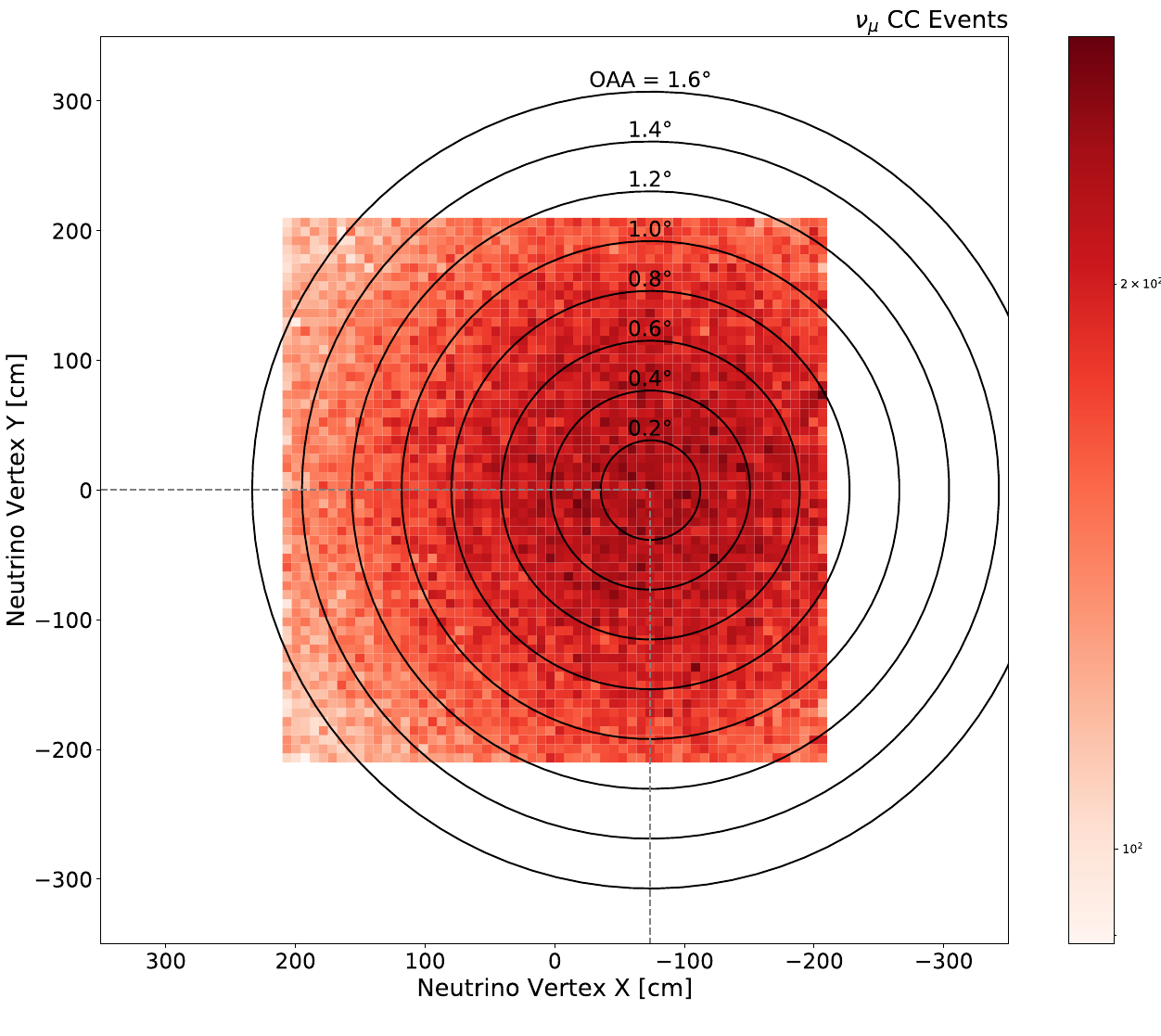}
    \caption{Left: Muon-neutrino fluxes at the SBND detector for different off-axis angles. Right: $\nu_\mu$ CC event rate at the front face of the SBND detector.}
    \label{fig:prism}
\end{figure}

SBND’s proximity to the beam target enables it to exploit a ``PRISM"-like feature -- the decrease in the peak energy of the neutrino spectrum and the reduction in the size of the high energy tail when the detection angle relative to the neutrino beam axis is increased~\cite{nuPRISM:2014mzw, DUNE:2021tad}. As shown in Fig.~\ref{fig:prism}, SBND can be divided into slices of off-axis angles ranging from $0.2^{\circ}$ to $1.6^{\circ}$, each exposed to a different neutrino flux. This feature introduces an additional handle for performing targeted neutrino cross-section measurements by enabling analysis of neutrino interactions across the range of off-axis angles, corresponding to fluxes with a mean energy shift of up to ${\sim} 200$~MeV. The SBND-PRISM ability of taking measurements in the same beam and with the same detector but with different neutrino flux spectra exposures by selecting events in different annular bins on the face of the detector provides unique constraints of systematic uncertainties, helps mitigate backgrounds, and expands the SBND physics potential.

\subsection{Advancing Liquid Argon TPC Technology}

SBND is a liquid argon TPC with two drift volumes separated by a central cathode plane and read out by wire planes (Anode Plane Assemblies) on the sides~\cite{SBND:2020scp}, in a configuration similar to the ProtoDUNE Horizontal Drift (NP04)~\cite{DUNE:2021hwx} at the CERN Neutrino Platform and the future DUNE Far Detector Horizontal Drift module~\cite{DUNE:2020txw}. SBND employs the same membrane-cryostat technology as DUNE, and SBND's cryostat and cryogenic proximity subsystems were provided by CERN as prototypes for the DUNE production~\cite{Geynisman:2017xpx}.

The readout of the drifted electrons is carried out using front-end electronics immersed in the liquid argon that share the first-stage components with the final DUNE design~\cite{Chen:2023buq}. 
In addition to the primary triggered readout, SBND also provides a second, parallel TPC readout stream that is capable of continuously reading out ionization signals from the detector, and would be capable of recording $\mathcal{O}(10)$ supernova neutrino interactions in the event of a galactic supernova burst during SBND’s operational lifetime. This TPC readout stream will also be used for the development and demonstration of TPC-based data selection and trigger strategies that will inform future LArTPCs such as DUNE~\cite{Karagiorgi:2019qzp}.

SBND features an advanced photon detection system equipped with Hamamatsu R5912-mod cryogenic PMTs~\cite{R5912datasheet} and X-ARAPUCAs~\cite{Machado:2018rfb}, a novel photodetector based on trapping photons in cavities equipped with SiPMs using dichroic filters and wavelength-shifters, which is also the choice for DUNE~\cite{DUNE:2020txw}. The X-ARAPUCA data from SBND will serve as the longest operation test (2024 -- 2027+) of this new photodetector technology in liquid argon before it is deployed in the DUNE Far Detectors, providing critical data on the performance stability, with the well-known PMTs installed side-by-side enabling direct comparisons. In addition, SBND's cathode is covered with TPB-coated reflective foils, a solution that has been considered for DUNE and dark matter detectors~\cite{DarkSide:2017wdu}.

SBND is a unique platform to continue developing the data analysis pipelines for LArTPC experiments exploiting the neutrino beam data, including the reconstruction of events with the TPC utilizing multiple paradigms (Pandora~\cite{MicroBooNE:2017xvs}, Wire-Cell~\cite{MicroBooNE:2021ojx}, deep learning methods~\cite{SBND:2020eho}), an advanced scintillation light reconstruction~\cite{SBND:2024vgn}, calibration techniques, and treatment of systematic uncertainties, in particular for a two-detector oscillation search.
In this regard, one of the most valuable aspects of SBND is the training of the next generation of LArTPC experts that will carry out the physics programs of this and future experiments.  

The European contributions to SBND are extensive, and their continuation is critical for the success of the experiment. 
In addition to the aforementioned contribution from CERN, the University of Bern provided and installed the UV laser calibration system that will be used to characterize the electric field in the TPC, and a cosmic-ray tagger (CRT) system made of scintillator strip planes surrounding the detector.
The UK designed, fabricated and tested major TPC components, namely half of the wire-readout planes, the cathode, and the novel reflective foil system that forms part of the photon detection system, which is integrated into the cathode. These precision components have additionally provided a pathway towards the industrial links needed for fabrication of DUNE components. There has been significant UK leadership in the assembly of the SBND detector, with involvement in the integration and installation of TPC and CRT components at Fermilab, as well as the subsequent commissioning phase. The UK is also contributing significantly to the development of LArTPC software tools, such as the Pandora reconstruction framework, the GENIE event generator, and the VALOR neutrino oscillation fitting tool. Spain has contributed to the readout of X-ARAPUCA photodetectors and the commissioning of the photon detection system, developing the scintillation light simulation and the ongoing reconstruction with data.

\section{Current Operations Plan and Long-Term Opportunities}\label{sec:operations_plans}

\begin{figure}
    \centering
    \includegraphics[width=0.32\linewidth]{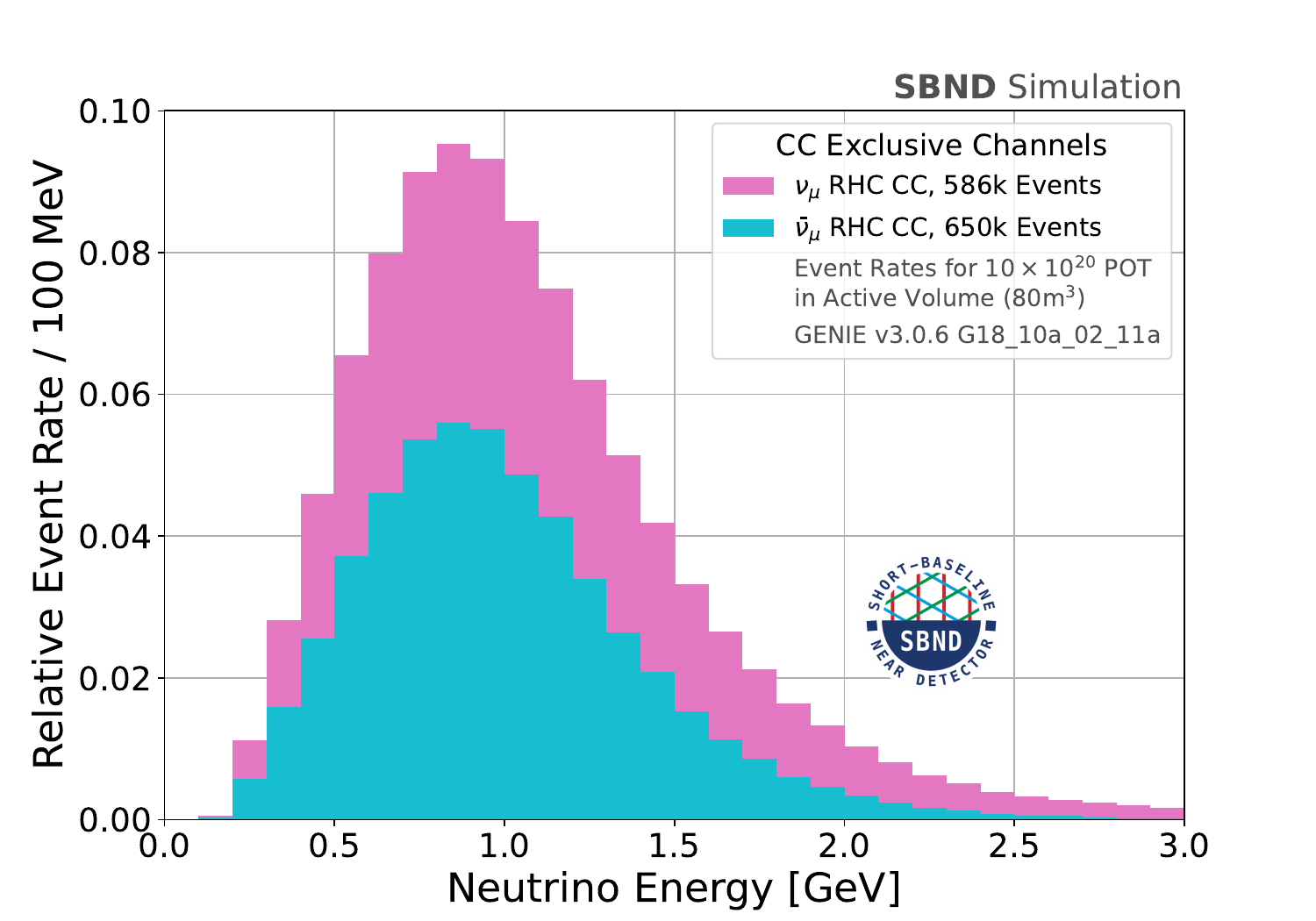}
    \includegraphics[width=0.32\linewidth]{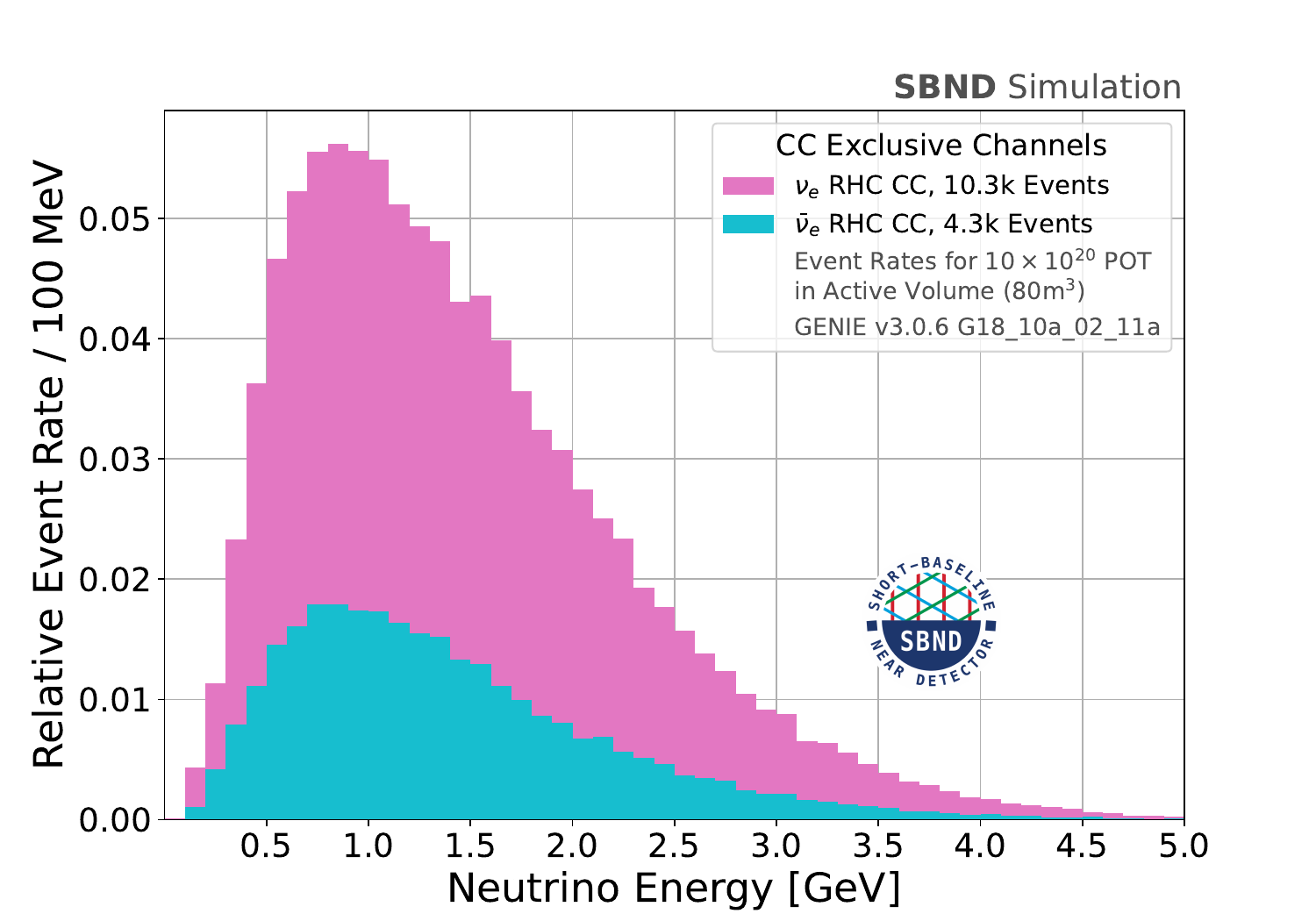}
    \includegraphics[width=0.33\linewidth]{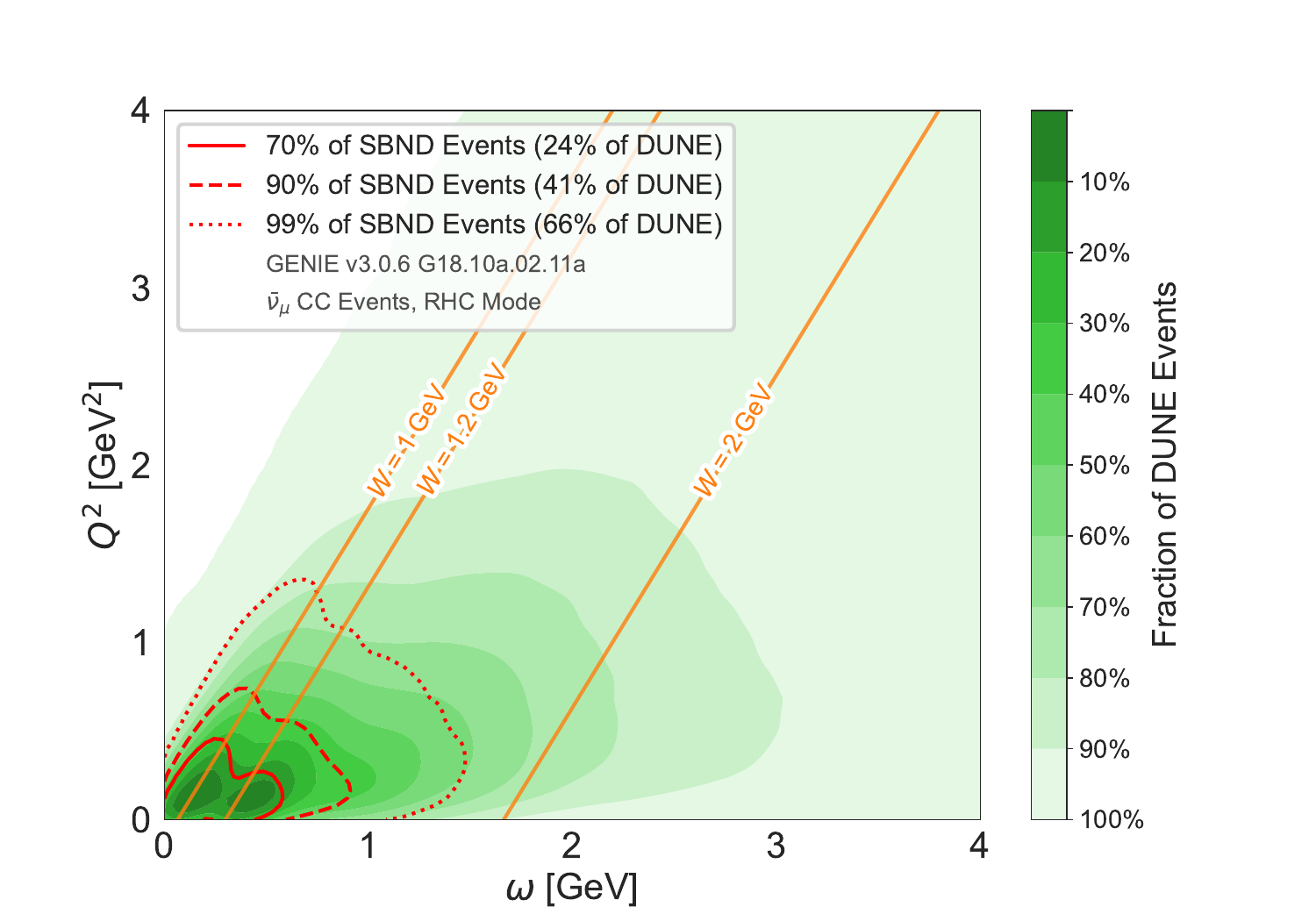}
    \caption{The expected CC interaction rates in SBND in the BNB antineutrino (RHC) mode for $\bar{\nu}_\mu$ \& $\nu_\mu$ (left) and $\bar{\nu}_e$ \& $\nu_e$ (middle), both plots are stacked. Note that electron (anti)neutrinos extend to much larger energies than muon (anti)neutrinos. Right: SBND $\bar{\nu}_\mu$ CC kinematic coverage overlaid on the DUNE $\bar{\nu}_\mu$ CC kinematic coverage.}
    \label{fig:antineutrino_event_rate}
\end{figure}

SBND began collecting BNB physics data in December 2024 and is expected to operate until the planned long accelerator shutdown at Fermilab scheduled for late 2027 or early 2028. This will result in a total exposure of approximately $10 \times 10^{20}$ protons on target yielding nearly 10 million neutrino interactions on argon across charged-current and neutral-current channels. This dataset will enable the physics program outlined above. For after the accelerator restart (2029+), opportunities are being explored to operate SBND with the BNB in antineutrino mode, which would quickly address the scarcity of antineutrino-argon scattering data, or in a dedicated beam-dump mode, enabling SBND to significantly enhance sensitivity to many new physics scenarios. Physics studies are currently under development, and operation past the long shutdown will be contingent upon host lab approval.

A precise understanding of antineutrino–argon interactions is essential, owing to potential differences in nuclear effects between neutrinos and antineutrinos driven by the isospin asymmetry of the argon nucleus, in the context of DUNE’s planned charge-parity (CP) violation studies. 
By reversing the horn current polarity, the BNB muon antineutrino fraction of flux increases from 6.9\% to 83.9\%, similar to past MiniBooNE operations in antineutrino mode~\cite{MiniBooNE:2008hfu, MiniBooNE:2013qnd}. Fig.~\ref{fig:antineutrino_event_rate} (left) shows the spectra of expected CC interactions for each neutrino species in the BNB antineutrino (reverse horn current, RHC) mode. Notably, electron (anti)neutrinos extend to significantly higher energies than muon (anti)neutrinos. This configuration enables SBND to collect the world’s largest sample of antineutrino–argon scattering events even within a relatively short operational period, providing a valuable dataset 
for DUNE. Fig.~\ref{fig:antineutrino_event_rate} (right) demonstrates that the SBND kinematic phase space overlaps substantially with that of DUNE. Leveraging SBND’s high-resolution LArTPC capabilities, these data would play a pivotal role in refining event generators and ensuring precise modeling of neutrino and antineutrino interactions for the next generation of experiments.
A dedicated antineutrino run may also be proven desirable for a more exhaustive search for oscillation signals within the context of extended light sterile neutrino oscillation models, e.g. 3+N. In this case, SBN could provide enhanced sensitivity to 3+2 and 3+3 short-baseline parameter space through multi-channel oscillation searches, as well as sensitivity to leptonic CP violation that may associated with an extended light sterile neutrino sector \cite{Cianci:2017okw,Acero:2022wqg}.

Another promising avenue would be to operate SBND in a beam dump configuration, either by steering the proton beam away from the beryllium target (as in MiniBooNE's off-target run~\cite{MiniBooNE:2017nqe, MiniBooNEDM:2018cxm}) or by installing a new, dedicated dense beam dump~\cite{Toups:2022knq}. This setup can suppress neutrino-induced backgrounds by a few orders of magnitude, significantly enhancing sensitivity to a broad range of new physics scenarios~\cite{Toups:2022knq}. Note that by the time post-long-shutdown data collection begins, the SBND detector is expected to be well-understood and well-calibrated, with significant progress made on systematics. This will enable a rapid turnaround of new results in antineutrino or beam-dump mode. Additionally, high-statistics neutrino interaction measurements in the neutrino (FHC) mode will help understand and constrain wrong-sign backgrounds, which will help mitigate the wrong-sign contamination observed in the antineutrino (RHC) mode, as seen in Fig.~\ref{fig:antineutrino_event_rate}.

\section{Summary}\label{sec:summary}

SBND is designed to address key questions in neutrino physics, including the search for eV-scale sterile neutrinos, precision neutrino-argon interaction measurements, and searches for physics beyond the Standard Model. Since its commissioning in 2024, SBND has begun collecting unprecedented amounts of neutrino data, significantly enhancing the sensitivity of the SBN Program and providing essential insights for the future flagship experiment DUNE. SBND has already accumulated the world’s largest neutrino-argon cross-section dataset and is expected to operate with a total exposure of around $10 \times 10^{20}$ protons on target amounting to nearly 10 million neutrino-argon interaction events. After the planned accelerator restart at Fermilab (2029+), opportunities are being explored to operate SBND in antineutrino mode, to address the scarcity of antineutrino–argon scattering data, or in a dedicated beam-dump mode to significantly enhance sensitivity to searches for new physics. With strong international collaboration -- particularly from European institutions -- SBND continues to be a vital platform for both technological advancements in liquid argon detectors and the training of the next generation of physicists. Continued support for SBND will be vital in ensuring the realization of the full physics potential of SBND/SBN and DUNE.


\acknowledgments
The SBND Collaboration acknowledges the generous support of the following organizations: 
the U.S. Department of Energy, Office of Science, Office of High Energy Physics; 
the U.S. National Science Foundation; 
the Science and Technology Facilities Council (STFC), part of United Kingdom Research and Innovation, The Royal Society of the United Kingdom, and the UK Research and Innovation (UKRI) Future Leaders Fellowship; 
the Swiss National Science Foundation; 
the Spanish Ministerio de Ciencia, Innovación e Universidades (MICIU/ AEI/ 10.13039/ 501100011033) under grants No PRE2019-090468, CNS2022-136022, RYC2022-036471-I, PID2023-147949NB-C51 \& C53; 
the European Union’s Horizon 2020 research and innovation programme under GA no 101004761 and the Marie Sklodowska\,-\,Curie grant agreements No 822185 and 101081478; 
the São Paulo Research Foundation 1098 (FAPESP), the National Council of Scientific and Technological Development (CNPq) and Ministry of  Science, Technology \& Innovations-MCTI of Brazil. 
We acknowledge Los Alamos National Laboratory for LDRD funding. 
This document was prepared by SBND Collaboration using the resources of the Fermi National Accelerator Laboratory (Fermilab), a U.S. Department of Energy, Office of Science, Office of High Energy Physics HEP User Facility. Fermilab is managed by FermiForward Discovery Group, LLC, acting under Contract No. 89243024CSC000002.

\bibliographystyle{apsrev4-1}
\bibliography{refs}

\begin{thebibliography}{46}%
\makeatletter
\providecommand \@ifxundefined [1]{%
 \@ifx{#1\undefined}
}%
\providecommand \@ifnum [1]{%
 \ifnum #1\expandafter \@firstoftwo
 \else \expandafter \@secondoftwo
 \fi
}%
\providecommand \@ifx [1]{%
 \ifx #1\expandafter \@firstoftwo
 \else \expandafter \@secondoftwo
 \fi
}%
\providecommand \natexlab [1]{#1}%
\providecommand \enquote  [1]{``#1''}%
\providecommand \bibnamefont  [1]{#1}%
\providecommand \bibfnamefont [1]{#1}%
\providecommand \citenamefont [1]{#1}%
\providecommand \href@noop [0]{\@secondoftwo}%
\providecommand \href [0]{\begingroup \@sanitize@url \@href}%
\providecommand \@href[1]{\@@startlink{#1}\@@href}%
\providecommand \@@href[1]{\endgroup#1\@@endlink}%
\providecommand \@sanitize@url [0]{\catcode `\\12\catcode `\$12\catcode `\&12\catcode `\#12\catcode `\^12\catcode `\_12\catcode `\%12\relax}%
\providecommand \@@startlink[1]{}%
\providecommand \@@endlink[0]{}%
\providecommand \url  [0]{\begingroup\@sanitize@url \@url }%
\providecommand \@url [1]{\endgroup\@href {#1}{\urlprefix }}%
\providecommand \urlprefix  [0]{URL }%
\providecommand \Eprint [0]{\href }%
\providecommand \doibase [0]{http://dx.doi.org/}%
\providecommand \selectlanguage [0]{\@gobble}%
\providecommand \bibinfo  [0]{\@secondoftwo}%
\providecommand \bibfield  [0]{\@secondoftwo}%
\providecommand \translation [1]{[#1]}%
\providecommand \BibitemOpen [0]{}%
\providecommand \bibitemStop [0]{}%
\providecommand \bibitemNoStop [0]{.\EOS\space}%
\providecommand \EOS [0]{\spacefactor3000\relax}%
\providecommand \BibitemShut  [1]{\csname bibitem#1\endcsname}%
\let\auto@bib@innerbib\@empty
\bibitem [{\citenamefont {Acciarri}\ \emph {et~al.}(2015)\citenamefont {Acciarri} \emph {et~al.}}]{MicroBooNE:2015bmn}%
  \BibitemOpen
  \bibfield  {author} {\bibinfo {author} {\bibfnamefont {R.}~\bibnamefont {Acciarri}} \emph {et~al.} (\bibinfo {collaboration} {MicroBooNE, LAr1-ND, ICARUS-WA104}),\ }\href@noop {} {\  (\bibinfo {year} {2015})},\ \Eprint {http://arxiv.org/abs/1503.01520} {arXiv:1503.01520 [physics.ins-det]} \BibitemShut {NoStop}%
\bibitem [{\citenamefont {Machado}\ \emph {et~al.}(2019)\citenamefont {Machado}, \citenamefont {Palamara},\ and\ \citenamefont {Schmitz}}]{Machado:2019oxb}%
  \BibitemOpen
  \bibfield  {author} {\bibinfo {author} {\bibfnamefont {P.~A.}\ \bibnamefont {Machado}}, \bibinfo {author} {\bibfnamefont {O.}~\bibnamefont {Palamara}}, \ and\ \bibinfo {author} {\bibfnamefont {D.~W.}\ \bibnamefont {Schmitz}},\ }\href {\doibase 10.1146/annurev-nucl-101917-020949} {\bibfield  {journal} {\bibinfo  {journal} {Ann. Rev. Nucl. Part. Sci.}\ }\textbf {\bibinfo {volume} {69}},\ \bibinfo {pages} {363} (\bibinfo {year} {2019})},\ \Eprint {http://arxiv.org/abs/1903.04608} {arXiv:1903.04608 [hep-ex]} \BibitemShut {NoStop}%
\bibitem [{\citenamefont {Abi}\ \emph {et~al.}(2020{\natexlab{a}})\citenamefont {Abi} \emph {et~al.}}]{DUNE:2020lwj}%
  \BibitemOpen
  \bibfield  {author} {\bibinfo {author} {\bibfnamefont {B.}~\bibnamefont {Abi}} \emph {et~al.} (\bibinfo {collaboration} {DUNE}),\ }\href {\doibase 10.1088/1748-0221/15/08/T08008} {\bibfield  {journal} {\bibinfo  {journal} {JINST}\ }\textbf {\bibinfo {volume} {15}},\ \bibinfo {pages} {T08008} (\bibinfo {year} {2020}{\natexlab{a}})},\ \Eprint {http://arxiv.org/abs/2002.02967} {arXiv:2002.02967 [physics.ins-det]} \BibitemShut {NoStop}%
\bibitem [{\citenamefont {Asai}\ \emph {et~al.}(2023)\citenamefont {Asai} \emph {et~al.}}]{P5:2023wyd}%
  \BibitemOpen
  \bibfield  {author} {\bibinfo {author} {\bibfnamefont {S.}~\bibnamefont {Asai}} \emph {et~al.} (\bibinfo {collaboration} {P5}),\ }\href {\doibase 10.2172/2368847} {\  (\bibinfo {year} {2023}),\ 10.2172/2368847},\ \Eprint {http://arxiv.org/abs/2407.19176} {arXiv:2407.19176 [hep-ex]} \BibitemShut {NoStop}%
\bibitem [{\citenamefont {Ellis}\ \emph {et~al.}(2019)\citenamefont {Ellis} \emph {et~al.}}]{EuropeanStrategyforParticlePhysicsPreparatoryGroup:2019qin}%
  \BibitemOpen
  \bibfield  {author} {\bibinfo {author} {\bibfnamefont {R.~K.}\ \bibnamefont {Ellis}} \emph {et~al.},\ }\href@noop {} {\  (\bibinfo {year} {2019})},\ \Eprint {http://arxiv.org/abs/1910.11775} {arXiv:1910.11775 [hep-ex]} \BibitemShut {NoStop}%
\bibitem [{\citenamefont {{European Strategy Group}}(2020)}]{CERN-ESU-015}%
  \BibitemOpen
  \bibfield  {author} {\bibinfo {author} {\bibnamefont {{European Strategy Group}}},\ }\href {\doibase 10.17181/CERN.JSC6.W89E} {\  (\bibinfo {year} {2020}),\ 10.17181/CERN.JSC6.W89E}\BibitemShut {NoStop}%
\bibitem [{\citenamefont {Furmanski}\ and\ \citenamefont {Hilgenberg}(2021)}]{Furmanski:2020smg}%
  \BibitemOpen
  \bibfield  {author} {\bibinfo {author} {\bibfnamefont {A.~P.}\ \bibnamefont {Furmanski}}\ and\ \bibinfo {author} {\bibfnamefont {C.}~\bibnamefont {Hilgenberg}},\ }\href {\doibase 10.1103/PhysRevD.103.112011} {\bibfield  {journal} {\bibinfo  {journal} {Phys. Rev. D}\ }\textbf {\bibinfo {volume} {103}},\ \bibinfo {pages} {112011} (\bibinfo {year} {2021})},\ \Eprint {http://arxiv.org/abs/2012.09788} {arXiv:2012.09788 [hep-ex]} \BibitemShut {NoStop}%
\bibitem [{\citenamefont {Dentler}\ \emph {et~al.}(2018)\citenamefont {Dentler}, \citenamefont {Hern\'andez-Cabezudo}, \citenamefont {Kopp}, \citenamefont {Machado}, \citenamefont {Maltoni}, \citenamefont {Martinez-Soler},\ and\ \citenamefont {Schwetz}}]{Dentler:2018sju}%
  \BibitemOpen
  \bibfield  {author} {\bibinfo {author} {\bibfnamefont {M.}~\bibnamefont {Dentler}}, \bibinfo {author} {\bibfnamefont {A.}~\bibnamefont {Hern\'andez-Cabezudo}}, \bibinfo {author} {\bibfnamefont {J.}~\bibnamefont {Kopp}}, \bibinfo {author} {\bibfnamefont {P.~A.~N.}\ \bibnamefont {Machado}}, \bibinfo {author} {\bibfnamefont {M.}~\bibnamefont {Maltoni}}, \bibinfo {author} {\bibfnamefont {I.}~\bibnamefont {Martinez-Soler}}, \ and\ \bibinfo {author} {\bibfnamefont {T.}~\bibnamefont {Schwetz}},\ }\href {\doibase 10.1007/JHEP08(2018)010} {\bibfield  {journal} {\bibinfo  {journal} {JHEP}\ }\textbf {\bibinfo {volume} {08}},\ \bibinfo {pages} {010} (\bibinfo {year} {2018})},\ \Eprint {http://arxiv.org/abs/1803.10661} {arXiv:1803.10661 [hep-ph]} \BibitemShut {NoStop}%
\bibitem [{\citenamefont {Mahn}\ \emph {et~al.}(2012)\citenamefont {Mahn} \emph {et~al.}}]{SciBooNE:2011qyf}%
  \BibitemOpen
  \bibfield  {author} {\bibinfo {author} {\bibfnamefont {K.~B.~M.}\ \bibnamefont {Mahn}} \emph {et~al.} (\bibinfo {collaboration} {SciBooNE, MiniBooNE}),\ }\href {\doibase 10.1103/PhysRevD.85.032007} {\bibfield  {journal} {\bibinfo  {journal} {Phys. Rev. D}\ }\textbf {\bibinfo {volume} {85}},\ \bibinfo {pages} {032007} (\bibinfo {year} {2012})},\ \Eprint {http://arxiv.org/abs/1106.5685} {arXiv:1106.5685 [hep-ex]} \BibitemShut {NoStop}%
\bibitem [{\citenamefont {Adamson}\ \emph {et~al.}(2019)\citenamefont {Adamson} \emph {et~al.}}]{MINOS:2017cae}%
  \BibitemOpen
  \bibfield  {author} {\bibinfo {author} {\bibfnamefont {P.}~\bibnamefont {Adamson}} \emph {et~al.} (\bibinfo {collaboration} {MINOS+}),\ }\href {\doibase 10.1103/PhysRevLett.122.091803} {\bibfield  {journal} {\bibinfo  {journal} {Phys. Rev. Lett.}\ }\textbf {\bibinfo {volume} {122}},\ \bibinfo {pages} {091803} (\bibinfo {year} {2019})},\ \Eprint {http://arxiv.org/abs/1710.06488} {arXiv:1710.06488 [hep-ex]} \BibitemShut {NoStop}%
\bibitem [{\citenamefont {Aartsen}\ \emph {et~al.}(2020)\citenamefont {Aartsen} \emph {et~al.}}]{IceCube:2020tka}%
  \BibitemOpen
  \bibfield  {author} {\bibinfo {author} {\bibfnamefont {M.~G.}\ \bibnamefont {Aartsen}} \emph {et~al.} (\bibinfo {collaboration} {IceCube}),\ }\href {\doibase 10.1103/PhysRevD.102.052009} {\bibfield  {journal} {\bibinfo  {journal} {Phys. Rev. D}\ }\textbf {\bibinfo {volume} {102}},\ \bibinfo {pages} {052009} (\bibinfo {year} {2020})},\ \Eprint {http://arxiv.org/abs/2005.12943} {arXiv:2005.12943 [hep-ex]} \BibitemShut {NoStop}%
\bibitem [{\citenamefont {Balantekin}\ \emph {et~al.}(2022)\citenamefont {Balantekin} \emph {et~al.}}]{Balantekin:2022jrq}%
  \BibitemOpen
  \bibfield  {author} {\bibinfo {author} {\bibfnamefont {A.~B.}\ \bibnamefont {Balantekin}} \emph {et~al.},\ }\href@noop {} {\  (\bibinfo {year} {2022})},\ \Eprint {http://arxiv.org/abs/2209.06872} {arXiv:2209.06872 [hep-ex]} \BibitemShut {NoStop}%
\bibitem [{\citenamefont {Ballett}\ \emph {et~al.}(2017)\citenamefont {Ballett}, \citenamefont {Pascoli},\ and\ \citenamefont {Ross-Lonergan}}]{Ballett:2016opr}%
  \BibitemOpen
  \bibfield  {author} {\bibinfo {author} {\bibfnamefont {P.}~\bibnamefont {Ballett}}, \bibinfo {author} {\bibfnamefont {S.}~\bibnamefont {Pascoli}}, \ and\ \bibinfo {author} {\bibfnamefont {M.}~\bibnamefont {Ross-Lonergan}},\ }\href {\doibase 10.1007/JHEP04(2017)102} {\bibfield  {journal} {\bibinfo  {journal} {JHEP}\ }\textbf {\bibinfo {volume} {04}},\ \bibinfo {pages} {102} (\bibinfo {year} {2017})},\ \Eprint {http://arxiv.org/abs/1610.08512} {arXiv:1610.08512 [hep-ph]} \BibitemShut {NoStop}%
\bibitem [{\citenamefont {Ballett}\ \emph {et~al.}(2020)\citenamefont {Ballett}, \citenamefont {Boschi},\ and\ \citenamefont {Pascoli}}]{Ballett:2019bgd}%
  \BibitemOpen
  \bibfield  {author} {\bibinfo {author} {\bibfnamefont {P.}~\bibnamefont {Ballett}}, \bibinfo {author} {\bibfnamefont {T.}~\bibnamefont {Boschi}}, \ and\ \bibinfo {author} {\bibfnamefont {S.}~\bibnamefont {Pascoli}},\ }\href {\doibase 10.1007/JHEP03(2020)111} {\bibfield  {journal} {\bibinfo  {journal} {JHEP}\ }\textbf {\bibinfo {volume} {03}},\ \bibinfo {pages} {111} (\bibinfo {year} {2020})},\ \Eprint {http://arxiv.org/abs/1905.00284} {arXiv:1905.00284 [hep-ph]} \BibitemShut {NoStop}%
\bibitem [{\citenamefont {Batell}\ \emph {et~al.}(2019)\citenamefont {Batell}, \citenamefont {Berger},\ and\ \citenamefont {Ismail}}]{Batell:2019nwo}%
  \BibitemOpen
  \bibfield  {author} {\bibinfo {author} {\bibfnamefont {B.}~\bibnamefont {Batell}}, \bibinfo {author} {\bibfnamefont {J.}~\bibnamefont {Berger}}, \ and\ \bibinfo {author} {\bibfnamefont {A.}~\bibnamefont {Ismail}},\ }\href {\doibase 10.1103/PhysRevD.100.115039} {\bibfield  {journal} {\bibinfo  {journal} {Phys. Rev. D}\ }\textbf {\bibinfo {volume} {100}},\ \bibinfo {pages} {115039} (\bibinfo {year} {2019})},\ \Eprint {http://arxiv.org/abs/1909.11670} {arXiv:1909.11670 [hep-ph]} \BibitemShut {NoStop}%
\bibitem [{\citenamefont {Berger}\ and\ \citenamefont {Putnam}(2024)}]{Berger:2024xqk}%
  \BibitemOpen
  \bibfield  {author} {\bibinfo {author} {\bibfnamefont {J.}~\bibnamefont {Berger}}\ and\ \bibinfo {author} {\bibfnamefont {G.}~\bibnamefont {Putnam}},\ }\href {\doibase 10.1103/PhysRevD.110.055035} {\bibfield  {journal} {\bibinfo  {journal} {Phys. Rev. D}\ }\textbf {\bibinfo {volume} {110}},\ \bibinfo {pages} {055035} (\bibinfo {year} {2024})},\ \Eprint {http://arxiv.org/abs/2405.18480} {arXiv:2405.18480 [hep-ph]} \BibitemShut {NoStop}%
\bibitem [{\citenamefont {Kelly}\ \emph {et~al.}(2021)\citenamefont {Kelly}, \citenamefont {Kumar},\ and\ \citenamefont {Liu}}]{Kelly:2020dda}%
  \BibitemOpen
  \bibfield  {author} {\bibinfo {author} {\bibfnamefont {K.~J.}\ \bibnamefont {Kelly}}, \bibinfo {author} {\bibfnamefont {S.}~\bibnamefont {Kumar}}, \ and\ \bibinfo {author} {\bibfnamefont {Z.}~\bibnamefont {Liu}},\ }\href {\doibase 10.1103/PhysRevD.103.095002} {\bibfield  {journal} {\bibinfo  {journal} {Phys. Rev. D}\ }\textbf {\bibinfo {volume} {103}},\ \bibinfo {pages} {095002} (\bibinfo {year} {2021})},\ \Eprint {http://arxiv.org/abs/2011.05995} {arXiv:2011.05995 [hep-ph]} \BibitemShut {NoStop}%
\bibitem [{\citenamefont {deNiverville}\ \emph {et~al.}(2017)\citenamefont {deNiverville}, \citenamefont {Chen}, \citenamefont {Pospelov},\ and\ \citenamefont {Ritz}}]{deNiverville:2016rqh}%
  \BibitemOpen
  \bibfield  {author} {\bibinfo {author} {\bibfnamefont {P.}~\bibnamefont {deNiverville}}, \bibinfo {author} {\bibfnamefont {C.-Y.}\ \bibnamefont {Chen}}, \bibinfo {author} {\bibfnamefont {M.}~\bibnamefont {Pospelov}}, \ and\ \bibinfo {author} {\bibfnamefont {A.}~\bibnamefont {Ritz}},\ }\href {\doibase 10.1103/PhysRevD.95.035006} {\bibfield  {journal} {\bibinfo  {journal} {Phys. Rev. D}\ }\textbf {\bibinfo {volume} {95}},\ \bibinfo {pages} {035006} (\bibinfo {year} {2017})},\ \Eprint {http://arxiv.org/abs/1609.01770} {arXiv:1609.01770 [hep-ph]} \BibitemShut {NoStop}%
\bibitem [{\citenamefont {Batell}\ \emph {et~al.}(2021)\citenamefont {Batell}, \citenamefont {Berger}, \citenamefont {Darm\'e},\ and\ \citenamefont {Frugiuele}}]{Batell:2021ooj}%
  \BibitemOpen
  \bibfield  {author} {\bibinfo {author} {\bibfnamefont {B.}~\bibnamefont {Batell}}, \bibinfo {author} {\bibfnamefont {J.}~\bibnamefont {Berger}}, \bibinfo {author} {\bibfnamefont {L.}~\bibnamefont {Darm\'e}}, \ and\ \bibinfo {author} {\bibfnamefont {C.}~\bibnamefont {Frugiuele}},\ }\href {\doibase 10.1103/PhysRevD.104.075026} {\bibfield  {journal} {\bibinfo  {journal} {Phys. Rev. D}\ }\textbf {\bibinfo {volume} {104}},\ \bibinfo {pages} {075026} (\bibinfo {year} {2021})},\ \Eprint {http://arxiv.org/abs/2106.04584} {arXiv:2106.04584 [hep-ph]} \BibitemShut {NoStop}%
\bibitem [{\citenamefont {Dutta}\ \emph {et~al.}(2022)\citenamefont {Dutta}, \citenamefont {Kim}, \citenamefont {Thompson}, \citenamefont {Thornton},\ and\ \citenamefont {Van~de Water}}]{Dutta:2021cip}%
  \BibitemOpen
  \bibfield  {author} {\bibinfo {author} {\bibfnamefont {B.}~\bibnamefont {Dutta}}, \bibinfo {author} {\bibfnamefont {D.}~\bibnamefont {Kim}}, \bibinfo {author} {\bibfnamefont {A.}~\bibnamefont {Thompson}}, \bibinfo {author} {\bibfnamefont {R.~T.}\ \bibnamefont {Thornton}}, \ and\ \bibinfo {author} {\bibfnamefont {R.~G.}\ \bibnamefont {Van~de Water}},\ }\href {\doibase 10.1103/PhysRevLett.129.111803} {\bibfield  {journal} {\bibinfo  {journal} {Phys. Rev. Lett.}\ }\textbf {\bibinfo {volume} {129}},\ \bibinfo {pages} {111803} (\bibinfo {year} {2022})},\ \Eprint {http://arxiv.org/abs/2110.11944} {arXiv:2110.11944 [hep-ph]} \BibitemShut {NoStop}%
\bibitem [{\citenamefont {Abdullahi}\ \emph {et~al.}(2025)\citenamefont {Abdullahi}, \citenamefont {Hoefken~Zink}, \citenamefont {Hostert}, \citenamefont {Massaro},\ and\ \citenamefont {Pascoli}}]{Abdullahi:2023ejc}%
  \BibitemOpen
  \bibfield  {author} {\bibinfo {author} {\bibfnamefont {A.~M.}\ \bibnamefont {Abdullahi}}, \bibinfo {author} {\bibfnamefont {J.}~\bibnamefont {Hoefken~Zink}}, \bibinfo {author} {\bibfnamefont {M.}~\bibnamefont {Hostert}}, \bibinfo {author} {\bibfnamefont {D.}~\bibnamefont {Massaro}}, \ and\ \bibinfo {author} {\bibfnamefont {S.}~\bibnamefont {Pascoli}},\ }\href {\doibase 10.1103/PhysRevD.111.035028} {\bibfield  {journal} {\bibinfo  {journal} {Phys. Rev. D}\ }\textbf {\bibinfo {volume} {111}},\ \bibinfo {pages} {035028} (\bibinfo {year} {2025})},\ \Eprint {http://arxiv.org/abs/2308.02543} {arXiv:2308.02543 [hep-ph]} \BibitemShut {NoStop}%
\bibitem [{\citenamefont {Batell}\ \emph {et~al.}(2023)\citenamefont {Batell}, \citenamefont {Huang},\ and\ \citenamefont {Kelly}}]{Batell:2023mdn}%
  \BibitemOpen
  \bibfield  {author} {\bibinfo {author} {\bibfnamefont {B.}~\bibnamefont {Batell}}, \bibinfo {author} {\bibfnamefont {W.}~\bibnamefont {Huang}}, \ and\ \bibinfo {author} {\bibfnamefont {K.~J.}\ \bibnamefont {Kelly}},\ }\href {\doibase 10.1007/JHEP08(2023)092} {\bibfield  {journal} {\bibinfo  {journal} {JHEP}\ }\textbf {\bibinfo {volume} {08}},\ \bibinfo {pages} {092} (\bibinfo {year} {2023})},\ \Eprint {http://arxiv.org/abs/2304.11189} {arXiv:2304.11189 [hep-ph]} \BibitemShut {NoStop}%
\bibitem [{\citenamefont {Abratenko}\ \emph {et~al.}(2024)\citenamefont {Abratenko} \emph {et~al.}}]{SBND:2024vgn}%
  \BibitemOpen
  \bibfield  {author} {\bibinfo {author} {\bibfnamefont {P.}~\bibnamefont {Abratenko}} \emph {et~al.} (\bibinfo {collaboration} {SBND}),\ }\href {\doibase 10.1140/epjc/s10052-024-13306-3} {\bibfield  {journal} {\bibinfo  {journal} {Eur. Phys. J. C}\ }\textbf {\bibinfo {volume} {84}},\ \bibinfo {pages} {1046} (\bibinfo {year} {2024})},\ \Eprint {http://arxiv.org/abs/2406.07514} {arXiv:2406.07514 [physics.ins-det]} \BibitemShut {NoStop}%
\bibitem [{\citenamefont {Catanesi}\ \emph {et~al.}(2007)\citenamefont {Catanesi} \emph {et~al.}}]{HARP:2007dqt}%
  \BibitemOpen
  \bibfield  {author} {\bibinfo {author} {\bibfnamefont {M.~G.}\ \bibnamefont {Catanesi}} \emph {et~al.} (\bibinfo {collaboration} {HARP}),\ }\href {\doibase 10.1140/epjc/s10052-007-0382-8} {\bibfield  {journal} {\bibinfo  {journal} {Eur. Phys. J. C}\ }\textbf {\bibinfo {volume} {52}},\ \bibinfo {pages} {29} (\bibinfo {year} {2007})},\ \Eprint {http://arxiv.org/abs/hep-ex/0702024} {arXiv:hep-ex/0702024} \BibitemShut {NoStop}%
\bibitem [{\citenamefont {Aguilar-Arevalo}\ \emph {et~al.}(2009)\citenamefont {Aguilar-Arevalo} \emph {et~al.}}]{MiniBooNE:2008hfu}%
  \BibitemOpen
  \bibfield  {author} {\bibinfo {author} {\bibfnamefont {A.~A.}\ \bibnamefont {Aguilar-Arevalo}} \emph {et~al.} (\bibinfo {collaboration} {MiniBooNE}),\ }\href {\doibase 10.1103/PhysRevD.79.072002} {\bibfield  {journal} {\bibinfo  {journal} {Phys. Rev. D}\ }\textbf {\bibinfo {volume} {79}},\ \bibinfo {pages} {072002} (\bibinfo {year} {2009})},\ \Eprint {http://arxiv.org/abs/0806.1449} {arXiv:0806.1449 [hep-ex]} \BibitemShut {NoStop}%
\bibitem [{\citenamefont {Nagai}(2021)}]{Nagai:2783037}%
  \BibitemOpen
  \bibfield  {author} {\bibinfo {author} {\bibfnamefont {Y.}~\bibnamefont {Nagai}} (\bibinfo {collaboration} {NA61/SHINE}),\ }\href {https://cds.cern.ch/record/2783037} {\emph {\bibinfo {title} {{Addendum to the NA61/SHINE Proposal: A Low-Energy Beamline at the SPS H2}}}},\ \bibinfo {type} {Tech. Rep.}\ (\bibinfo  {institution} {CERN},\ \bibinfo {address} {Geneva},\ \bibinfo {year} {2021})\BibitemShut {NoStop}%
\bibitem [{\citenamefont {Bhadra}\ \emph {et~al.}(2014)\citenamefont {Bhadra} \emph {et~al.}}]{nuPRISM:2014mzw}%
  \BibitemOpen
  \bibfield  {author} {\bibinfo {author} {\bibfnamefont {S.}~\bibnamefont {Bhadra}} \emph {et~al.} (\bibinfo {collaboration} {nuPRISM}),\ }\href@noop {} {\  (\bibinfo {year} {2014})},\ \Eprint {http://arxiv.org/abs/1412.3086} {arXiv:1412.3086 [physics.ins-det]} \BibitemShut {NoStop}%
\bibitem [{\citenamefont {Abud}\ \emph {et~al.}(2021)\citenamefont {Abud} \emph {et~al.}}]{DUNE:2021tad}%
  \BibitemOpen
  \bibfield  {author} {\bibinfo {author} {\bibfnamefont {A.~A.}\ \bibnamefont {Abud}} \emph {et~al.} (\bibinfo {collaboration} {DUNE}),\ }\href {\doibase 10.3390/instruments5040031} {\bibfield  {journal} {\bibinfo  {journal} {Instruments}\ }\textbf {\bibinfo {volume} {5}},\ \bibinfo {pages} {31} (\bibinfo {year} {2021})},\ \Eprint {http://arxiv.org/abs/2103.13910} {arXiv:2103.13910 [physics.ins-det]} \BibitemShut {NoStop}%
\bibitem [{\citenamefont {Acciarri}\ \emph {et~al.}(2020)\citenamefont {Acciarri} \emph {et~al.}}]{SBND:2020scp}%
  \BibitemOpen
  \bibfield  {author} {\bibinfo {author} {\bibfnamefont {R.}~\bibnamefont {Acciarri}} \emph {et~al.} (\bibinfo {collaboration} {SBND}),\ }\href {\doibase 10.1088/1748-0221/15/06/P06033} {\bibfield  {journal} {\bibinfo  {journal} {JINST}\ }\textbf {\bibinfo {volume} {15}},\ \bibinfo {pages} {P06033} (\bibinfo {year} {2020})},\ \Eprint {http://arxiv.org/abs/2002.08424} {arXiv:2002.08424 [physics.ins-det]} \BibitemShut {NoStop}%
\bibitem [{\citenamefont {Abud}\ \emph {et~al.}(2022)\citenamefont {Abud} \emph {et~al.}}]{DUNE:2021hwx}%
  \BibitemOpen
  \bibfield  {author} {\bibinfo {author} {\bibfnamefont {A.~A.}\ \bibnamefont {Abud}} \emph {et~al.} (\bibinfo {collaboration} {DUNE}),\ }\href {\doibase 10.1088/1748-0221/17/01/P01005} {\bibfield  {journal} {\bibinfo  {journal} {JINST}\ }\textbf {\bibinfo {volume} {17}},\ \bibinfo {pages} {P01005} (\bibinfo {year} {2022})},\ \Eprint {http://arxiv.org/abs/2108.01902} {arXiv:2108.01902 [physics.ins-det]} \BibitemShut {NoStop}%
\bibitem [{\citenamefont {Abi}\ \emph {et~al.}(2020{\natexlab{b}})\citenamefont {Abi} \emph {et~al.}}]{DUNE:2020txw}%
  \BibitemOpen
  \bibfield  {author} {\bibinfo {author} {\bibfnamefont {B.}~\bibnamefont {Abi}} \emph {et~al.} (\bibinfo {collaboration} {DUNE}),\ }\href {\doibase 10.1088/1748-0221/15/08/T08010} {\bibfield  {journal} {\bibinfo  {journal} {JINST}\ }\textbf {\bibinfo {volume} {15}},\ \bibinfo {pages} {T08010} (\bibinfo {year} {2020}{\natexlab{b}})},\ \Eprint {http://arxiv.org/abs/2002.03010} {arXiv:2002.03010 [physics.ins-det]} \BibitemShut {NoStop}%
\bibitem [{\citenamefont {Geynisman}\ \emph {et~al.}(2017)\citenamefont {Geynisman} \emph {et~al.}}]{Geynisman:2017xpx}%
  \BibitemOpen
  \bibfield  {author} {\bibinfo {author} {\bibfnamefont {M.}~\bibnamefont {Geynisman}} \emph {et~al.},\ }\href {\doibase 10.1088/1757-899X/278/1/012119} {\bibfield  {journal} {\bibinfo  {journal} {IOP Conf. Ser. Mater. Sci. Eng.}\ }\textbf {\bibinfo {volume} {278}},\ \bibinfo {pages} {012119} (\bibinfo {year} {2017})}\BibitemShut {NoStop}%
\bibitem [{\citenamefont {Chen}\ and\ \citenamefont {Radeka}(2023)}]{Chen:2023buq}%
  \BibitemOpen
  \bibfield  {author} {\bibinfo {author} {\bibfnamefont {H.}~\bibnamefont {Chen}}\ and\ \bibinfo {author} {\bibfnamefont {V.}~\bibnamefont {Radeka}},\ }\href {\doibase 10.1016/j.nima.2022.167571} {\bibfield  {journal} {\bibinfo  {journal} {Nucl. Instrum. Meth. A}\ }\textbf {\bibinfo {volume} {1045}},\ \bibinfo {pages} {167571} (\bibinfo {year} {2023})}\BibitemShut {NoStop}%
\bibitem [{\citenamefont {Karagiorgi}(2019)}]{Karagiorgi:2019qzp}%
  \BibitemOpen
  \bibfield  {author} {\bibinfo {author} {\bibfnamefont {G.}~\bibnamefont {Karagiorgi}} (\bibinfo {collaboration} {SBND}),\ }in\ \href@noop {} {\emph {\bibinfo {booktitle} {{Meeting of the Division of Particles and Fields of the American Physical Society}}}}\ (\bibinfo {year} {2019})\ \Eprint {http://arxiv.org/abs/1910.08218} {arXiv:1910.08218 [physics.ins-det]} \BibitemShut {NoStop}%
\bibitem [{R59()}]{R5912datasheet}%
  \BibitemOpen
  \href@noop {} {\enquote {\bibinfo {title} {{HAMAMATSU} {D}atasheet - {R}5912},}\ }\bibinfo {howpublished} {\url{https://www.hamamatsu.com/resources/pdf/etd/LARGE_AREA_PMT_TPMH1376E.pdf}},\ \bibinfo {note} {accessed: September 2021}\BibitemShut {NoStop}%
\bibitem [{\citenamefont {Machado}\ \emph {et~al.}(2018)\citenamefont {Machado}, \citenamefont {Segreto}, \citenamefont {Warner}, \citenamefont {Fauth}, \citenamefont {Gelli}, \citenamefont {Maximo}, \citenamefont {Pizolatti}, \citenamefont {Paulucci},\ and\ \citenamefont {Marinho}}]{Machado:2018rfb}%
  \BibitemOpen
  \bibfield  {author} {\bibinfo {author} {\bibfnamefont {A.~A.}\ \bibnamefont {Machado}}, \bibinfo {author} {\bibfnamefont {E.}~\bibnamefont {Segreto}}, \bibinfo {author} {\bibfnamefont {D.}~\bibnamefont {Warner}}, \bibinfo {author} {\bibfnamefont {A.}~\bibnamefont {Fauth}}, \bibinfo {author} {\bibfnamefont {B.}~\bibnamefont {Gelli}}, \bibinfo {author} {\bibfnamefont {R.}~\bibnamefont {Maximo}}, \bibinfo {author} {\bibfnamefont {A.}~\bibnamefont {Pizolatti}}, \bibinfo {author} {\bibfnamefont {L.}~\bibnamefont {Paulucci}}, \ and\ \bibinfo {author} {\bibfnamefont {F.}~\bibnamefont {Marinho}}\ }(\bibinfo {year} {2018})\ \Eprint {http://arxiv.org/abs/1804.01407} {arXiv:1804.01407 [physics.ins-det]} \BibitemShut {NoStop}%
\bibitem [{\citenamefont {Agnes}\ \emph {et~al.}(2017)\citenamefont {Agnes} \emph {et~al.}}]{DarkSide:2017wdu}%
  \BibitemOpen
  \bibfield  {author} {\bibinfo {author} {\bibfnamefont {P.}~\bibnamefont {Agnes}} \emph {et~al.} (\bibinfo {collaboration} {DarkSide}),\ }\href {\doibase 10.1088/1748-0221/12/10/P10015} {\bibfield  {journal} {\bibinfo  {journal} {JINST}\ }\textbf {\bibinfo {volume} {12}},\ \bibinfo {pages} {P10015} (\bibinfo {year} {2017})},\ \Eprint {http://arxiv.org/abs/1707.05630} {arXiv:1707.05630 [physics.ins-det]} \BibitemShut {NoStop}%
\bibitem [{\citenamefont {Acciarri}\ \emph {et~al.}(2018)\citenamefont {Acciarri} \emph {et~al.}}]{MicroBooNE:2017xvs}%
  \BibitemOpen
  \bibfield  {author} {\bibinfo {author} {\bibfnamefont {R.}~\bibnamefont {Acciarri}} \emph {et~al.} (\bibinfo {collaboration} {MicroBooNE}),\ }\href {\doibase 10.1140/epjc/s10052-017-5481-6} {\bibfield  {journal} {\bibinfo  {journal} {Eur. Phys. J. C}\ }\textbf {\bibinfo {volume} {78}},\ \bibinfo {pages} {82} (\bibinfo {year} {2018})},\ \Eprint {http://arxiv.org/abs/1708.03135} {arXiv:1708.03135 [hep-ex]} \BibitemShut {NoStop}%
\bibitem [{\citenamefont {Abratenko}\ \emph {et~al.}(2022)\citenamefont {Abratenko} \emph {et~al.}}]{MicroBooNE:2021ojx}%
  \BibitemOpen
  \bibfield  {author} {\bibinfo {author} {\bibfnamefont {P.}~\bibnamefont {Abratenko}} \emph {et~al.} (\bibinfo {collaboration} {MicroBooNE}),\ }\href {\doibase 10.1088/1748-0221/17/01/P01037} {\bibfield  {journal} {\bibinfo  {journal} {JINST}\ }\textbf {\bibinfo {volume} {17}},\ \bibinfo {pages} {P01037} (\bibinfo {year} {2022})},\ \Eprint {http://arxiv.org/abs/2110.13961} {arXiv:2110.13961 [physics.ins-det]} \BibitemShut {NoStop}%
\bibitem [{\citenamefont {Acciarri}\ \emph {et~al.}(2021)\citenamefont {Acciarri} \emph {et~al.}}]{SBND:2020eho}%
  \BibitemOpen
  \bibfield  {author} {\bibinfo {author} {\bibfnamefont {R.}~\bibnamefont {Acciarri}} \emph {et~al.} (\bibinfo {collaboration} {SBND}),\ }\href {\doibase 10.3389/frai.2021.649917} {\bibfield  {journal} {\bibinfo  {journal} {Front. Artif. Intell.}\ }\textbf {\bibinfo {volume} {4}},\ \bibinfo {pages} {649917} (\bibinfo {year} {2021})},\ \Eprint {http://arxiv.org/abs/2012.01301} {arXiv:2012.01301 [physics.data-an]} \BibitemShut {NoStop}%
\bibitem [{\citenamefont {Aguilar-Arevalo}\ \emph {et~al.}(2013)\citenamefont {Aguilar-Arevalo} \emph {et~al.}}]{MiniBooNE:2013qnd}%
  \BibitemOpen
  \bibfield  {author} {\bibinfo {author} {\bibfnamefont {A.~A.}\ \bibnamefont {Aguilar-Arevalo}} \emph {et~al.} (\bibinfo {collaboration} {MiniBooNE}),\ }\href {\doibase 10.1103/PhysRevD.88.032001} {\bibfield  {journal} {\bibinfo  {journal} {Phys. Rev. D}\ }\textbf {\bibinfo {volume} {88}},\ \bibinfo {pages} {032001} (\bibinfo {year} {2013})},\ \Eprint {http://arxiv.org/abs/1301.7067} {arXiv:1301.7067 [hep-ex]} \BibitemShut {NoStop}%
\bibitem [{\citenamefont {Cianci}\ \emph {et~al.}(2017)\citenamefont {Cianci}, \citenamefont {Furmanski}, \citenamefont {Karagiorgi},\ and\ \citenamefont {Ross-Lonergan}}]{Cianci:2017okw}%
  \BibitemOpen
  \bibfield  {author} {\bibinfo {author} {\bibfnamefont {D.}~\bibnamefont {Cianci}}, \bibinfo {author} {\bibfnamefont {A.}~\bibnamefont {Furmanski}}, \bibinfo {author} {\bibfnamefont {G.}~\bibnamefont {Karagiorgi}}, \ and\ \bibinfo {author} {\bibfnamefont {M.}~\bibnamefont {Ross-Lonergan}},\ }\href {\doibase 10.1103/PhysRevD.96.055001} {\bibfield  {journal} {\bibinfo  {journal} {Phys. Rev. D}\ }\textbf {\bibinfo {volume} {96}},\ \bibinfo {pages} {055001} (\bibinfo {year} {2017})},\ \Eprint {http://arxiv.org/abs/1702.01758} {arXiv:1702.01758 [hep-ph]} \BibitemShut {NoStop}%
\bibitem [{\citenamefont {Acero}\ \emph {et~al.}(2024)\citenamefont {Acero} \emph {et~al.}}]{Acero:2022wqg}%
  \BibitemOpen
  \bibfield  {author} {\bibinfo {author} {\bibfnamefont {M.~A.}\ \bibnamefont {Acero}} \emph {et~al.},\ }\href {\doibase 10.1088/1361-6471/ad307f} {\bibfield  {journal} {\bibinfo  {journal} {J. Phys. G}\ }\textbf {\bibinfo {volume} {51}},\ \bibinfo {pages} {120501} (\bibinfo {year} {2024})},\ \Eprint {http://arxiv.org/abs/2203.07323} {arXiv:2203.07323 [hep-ex]} \BibitemShut {NoStop}%
\bibitem [{\citenamefont {Aguilar-Arevalo}\ \emph {et~al.}(2017)\citenamefont {Aguilar-Arevalo} \emph {et~al.}}]{MiniBooNE:2017nqe}%
  \BibitemOpen
  \bibfield  {author} {\bibinfo {author} {\bibfnamefont {A.~A.}\ \bibnamefont {Aguilar-Arevalo}} \emph {et~al.} (\bibinfo {collaboration} {MiniBooNE}),\ }\href {\doibase 10.1103/PhysRevLett.118.221803} {\bibfield  {journal} {\bibinfo  {journal} {Phys. Rev. Lett.}\ }\textbf {\bibinfo {volume} {118}},\ \bibinfo {pages} {221803} (\bibinfo {year} {2017})},\ \Eprint {http://arxiv.org/abs/1702.02688} {arXiv:1702.02688 [hep-ex]} \BibitemShut {NoStop}%
\bibitem [{\citenamefont {Aguilar-Arevalo}\ \emph {et~al.}(2018)\citenamefont {Aguilar-Arevalo} \emph {et~al.}}]{MiniBooNEDM:2018cxm}%
  \BibitemOpen
  \bibfield  {author} {\bibinfo {author} {\bibfnamefont {A.~A.}\ \bibnamefont {Aguilar-Arevalo}} \emph {et~al.} (\bibinfo {collaboration} {MiniBooNE DM}),\ }\href {\doibase 10.1103/PhysRevD.98.112004} {\bibfield  {journal} {\bibinfo  {journal} {Phys. Rev. D}\ }\textbf {\bibinfo {volume} {98}},\ \bibinfo {pages} {112004} (\bibinfo {year} {2018})},\ \Eprint {http://arxiv.org/abs/1807.06137} {arXiv:1807.06137 [hep-ex]} \BibitemShut {NoStop}%
\bibitem [{\citenamefont {Toups}\ \emph {et~al.}(2022)\citenamefont {Toups} \emph {et~al.}}]{Toups:2022knq}%
  \BibitemOpen
  \bibfield  {author} {\bibinfo {author} {\bibfnamefont {M.}~\bibnamefont {Toups}} \emph {et~al.},\ }in\ \href@noop {} {\emph {\bibinfo {booktitle} {{Snowmass 2021}}}}\ (\bibinfo {year} {2022})\ \Eprint {http://arxiv.org/abs/2203.08102} {arXiv:2203.08102 [hep-ex]} \BibitemShut {NoStop}%
\end{thebibliography}%

\end{document}